\documentclass[%
 reprint,
 superscriptaddress,
 preprintnumbers,
 nofootinbib,
 amsmath,amssymb,
 aps,
 prd,
 longbibliography,
 showkeys,
]{revtex4-1}

\usepackage{graphicx}% Include figure files
\usepackage{dcolumn}% Align table columns on decimal point
\usepackage{bm}% bold math
\usepackage{float}
\usepackage{hyperref}% add hypertext capabilities
\hypersetup{colorlinks=true, linkcolor=blue, filecolor=magenta,  urlcolor=blue, citecolor=cyan,}
\usepackage[dvipsnames,table]{xcolor}
\def\half{{\textstyle{1\over2}}}
\def\AA{\bm A}
\def\BB{\bm B}
\def\EE{\bm E}
\def\ee{\bm e}
\def\kk{\bm k}
\def\xx{\bm x}
\def\kap{\kappa}
\def\sig{\sigma}
\def\gam{\gamma}

\def\p{\partial}
\def\munu{{\mu\nu}}

\def\del{\nabla}
\def\Del{\bm\nabla}
\def\DDel{\bm\nabla^2}

\def\AAA{\mathcal{A}}
\def\EEE{\mathcal{E}}
\def\LLL{\mathcal{L}}
\def\OOO{\mathcal{O}}
\def\PPP{\mathcal{P}}
\def\GN{G_{\rm N}}

\def\nlin{{\rm nlin}}
\def\hel{{\rm hel}}
\def\nhel{{\rm nhel}}
\def\EEEM{\mathcal{E}_{\rm EM}}
\def\EEGW{\mathcal{E}_{\rm GW}}

\def\HGW{H_{\rm GW}}
\def\PPGW{\mathcal{P}_{\rm GW}}

\def\kNy{k_{\rm Ny}}
\def\EGW{E_{\rm GW}}

\def\GW{{\rm GW}}

\def\OmGW{\Omega_{\rm GW}}
\def\XiGW{\Xi_{\rm GW}}
\def\gr{g_{\rm r}}
\def\Tr{T_{\rm r}}
\def\kB{k_{\rm B}}
\def\Im{{\rm Im}}
\def\crit{{\rm crit}}

\def\MeV{{\rm \, MeV}}
\def\GeV{{\rm \, GeV}}
\def\TeV{{\rm \, TeV}}
\def\Hz{{\rm \, Hz}}
\def\km{{\rm \, km}}
\def\s{{\rm \, s}}
\def\Mpc{{\rm \, Mpc}}
\def\WKB{{\rm \, WKB}}
\def\EM{{\rm EM}} 
\def\rad{{\rm rad}}
\def\hel{{\rm hel}}
\def\nhel{{\rm nhel}}

\newcommand{\dd}{{\rm \, d}}
\newcommand{\bra}[1]{\langle #1\rangle}
\newcommand{\Eq}[1]{Eq.~(\ref{#1})}
\newcommand{\Eqs}[2]{Eqs.~(\ref{#1}) and (\ref{#2})}
\newcommand{\Fig}[1]{Fig.~\ref{#1}}
\newcommand{\Tab}[1]{Table~\ref{#1}}
\newcommand{\Sec}[1]{Sec.~\ref{#1}}
\newcommand{\Secs}[2]{Secs.~\ref{#1} and \ref{#2}}
\newcommand{\blue}[1]{\textcolor{blue}{#1}}
\newcommand{\green}[1]{\textcolor{ForestGreen}{#1}}
\newcommand{\orange}[1]{\textcolor{orange}{#1}}
\newcommand{\red}[1]{\textcolor{red}{#1}}
\newcommand{\purple}[1]{\textcolor{purple}{#1}}

\begin{document}

\preprint{NORDITA-2021-080}

\title{Leading-order nonlinear gravitational waves from reheating magnetogeneses}

\author{Yutong He}
\email{yutong.he@su.se}
\affiliation{Nordita, KTH Royal Institute of Technology and Stockholm University,Hannes Alfv\'ens v\"ag 12, 10691 Stockholm, Sweden}
\affiliation{Department of Astronomy, AlbaNova University Center, Stockholm University, 10691 Stockholm, Sweden}

\author{Alberto Roper Pol}
\email{roperpol@apc.in2p3.fr}
\affiliation{Universit\'e de Paris, CNRS, Astrophysique et Cosmologie, Paris, F-75013, France}
\affiliation{School of Natural Sciences and Medicine, Ilia State University, 3-5 Cholokashvili Ave, Tbilisi, GE-0194, Georgia}

\author{Axel Brandenburg}
\email{brandenb@nordita.org}
\affiliation{Nordita, KTH Royal Institute of Technology and Stockholm University,Hannes Alfv\'ens v\"ag 12, 10691 Stockholm, Sweden}
\affiliation{Department of Astronomy, AlbaNova University Center, Stockholm University, 10691 Stockholm, Sweden}
\affiliation{School of Natural Sciences and Medicine, Ilia State University, 3-5 Cholokashvili Ave, Tbilisi, GE-0194, Georgia}
\affiliation{McWilliams Center for Cosmology and Department of Physics, Carnegie Mellon University, 5000 Forbes Ave, Pittsburgh, PA 15213, USA}

\date{\today}

\begin{abstract}
We study the leading-order nonlinear gravitational waves (GWs) produced
by an electromagnetic (EM) stress in reheating
magnetogenesis scenarios.
Both nonhelical and helical magnetic fields are considered.
By numerically solving the linear and leading-order nonlinear GW
equations, we find that the GW energy from the latter is usually larger.
We compare their differences in terms of the GW spectrum
and parameterize the GW energy
difference due to the nonlinear term, $\Delta\EEGW$, in terms of EM energy
$\EEEM$ as $\Delta\EEGW=(\tilde p\EEEM/k_*)^3$, where $k_*$ is the characteristic wave number,
$\tilde p=0.84$ and $0.88$ are found in
the nonhelical and helical cases, respectively, with reheating around the QCD energy
scale, while $\tilde p=0.45$ is found at the electroweak energy scale.
We also compare the polarization spectrum of the linear and
nonlinear cases and find that adding the nonlinear term usually yields
a decrease in the polarization that is
proportional to the EM energy density.
We parameterize the fractional polarization suppression as
$|\Delta \PPGW/\PPGW|=\tilde r \EEEM/k_*$ and find $\tilde r = 1.2 \times 10^{-1}$, $7.2 \times 10^{-4}$,
and $3.2 \times 10^{-2}$ for the helical cases with reheating temperatures $\Tr =
300 \TeV$, $8 \GeV$, and $120 \MeV$, respectively.
Prospects of observation by pulsar timing arrays, space-based interferometers, and other 
novel detection proposals are also discussed.
\end{abstract}


\maketitle

%\tableofcontents

\section{Introduction}

The recent detections of gravitational wave (GW) events produced by the collision of compact binary objects by the LIGO-Virgo collaboration 
mark the beginning of the rich field of 
GW astronomy \cite{LIGOScientific:2016aoc,LIGOScientific:2016sjg,LIGOScientific:2017vwq}.
Over 50 events have been detected so far by Advanced LIGO and
Advanced
Virgo,
which are collected in the GW
transient catalogs \cite{LIGOScientific:2018mvr,LIGOScientific:2020ibl}.
These events yield propagating oscillatory strains of the metric tensor, which were
predicted by linearized general relativity (GR).
On the other hand,
a different type of perturbations in the spacetime metric yielding non-oscillatory strains
have been predicted by nonlinear GR, although they have
not yet been detected.
This effect, known as GW memory, is manifested by a permanent displacement of freely-falling test masses after the passage of GWs
\cite{Zeldovich:1974gvh,Turner78,Epstein78,Braginskii:1987,Christodoulou91,Blanchet+92,Thorne92}. 
It is a consequence of the facts that GR admits highly-degenerate
Bondi-Metzner-Sachs (BMS) vacua \cite{Bondi:1962px,Sachs:1962wk} 
and that the initial and final metrics, although both flat, differ by a BMS supertranslation induced by the passing GWs.
Theoretically, it has been shown that the GW memory effect is equivalent to the temporal Fourier transform of the Weinberg soft graviton
theorem \cite{Strominger+14,He+14}, 
which is a universal formula relating to each other two scattering matrices that differ by a zero-energy (soft) graviton \cite{Weinberg:1965nx}.

Recently, the memory effect has seen a pick-up in interest
in the GW community extensively due to the expected improvement in
the coming years in
sensitivity by upgraded GW detectors.
In particular, the advanced LIGO-Virgo-KAGRA network
will be able to detect an ensemble of memory signals
from binary mergers \cite{Lasky+16,McNeill+17,Hubner+19,Boersma+20}.
On the other hand, space-based
interferometers like the Laser Interferometer Space Antenna
(LISA), planned to be launched in 2034 \cite{LISA:2017pwj}, will be capable of
storing the permanent displacements since the test masses are
actually free-falling, unlike in ground-based detectors \cite{Favata09}.
It is expected that LISA will detect compact binaries
memory signals individually \cite{Favata09,Islo+19,Yang+18}. 
In the lower-frequency end, pulsar timing arrays (PTAs)
are also considered as potential memory detectors and
certain constraints have been put on the detectable signals
\cite{Wang+14,NANOGrav15,Madison+17,NANOGrav19}.
Astrophysical sources of the memory effect have been studied
in binary black holes
\cite{Favata09,Pollney+10,Khera+20,Ebersold+20,Islo+19},
supernova neutrinos
\cite{Turner78,Epstein78,Mukhopadhyay+21}, and gamma-ray
burst jets \cite{Burrows:1995bb,Segalis:2001ns,Sago:2004pn,Hiramatsu:2005jn}.
As for cosmological sources, cosmic
strings have been considered \cite{Aurrekoetxea+20,Jenkins+21}. 
In particular, the memory strain from compact binary objects has been
expected to be only one order of magnitude weaker than the standard GW
strain \cite{Favata09,Khera+20}.
As a test of gravity theories, the memory effect has been studied as a
result of massive gravity \cite{Kilicarslan+18}, scalar-tensor theories
\cite{Hou20,Hou+20,Koyama20}, as well as in the strong-field regime
\cite{Shore18}.
In addition, novel phenomena, different from the conventional displacement memory effect, have been proposed.
These are the velocity
\cite{Grishchuk+89,Zhang+17,Zhang+18,Divakarla+21} and spin memory effect \cite{Pasterski+15,Nichols17,Mitman+20,Tahura+21},
where the freely-falling particles would acquire permanent relative velocities and
spins, respectively, after the passage of GWs.

However, despite the plethora of literature on GW memory effect to date, to the best of our knowledge, 
none has explored the effect with continuous sources on cosmological scales, which are ubiquitous in the early universe.
With the recent advance of numerical simulations, there has been
significant progress in calculating the GW backgrounds (GWBs) from distributed
sources such as homogeneous hydrodynamic and hydromagnetic turbulence
in the early universe
\cite{Pol+18,RoperPol:2019wvy}.
We are interested in studying the leading-order nonlinear term producing
gravitational radiation, for which we require strong sources.
For this reason, we consider in this work the production of GWs by
reheating magnetogeneses, which yield electromagnetic (EM) strengths
that depend only on the initial field and can grow several orders
of magnitude \cite{Brandenburg:2021pdv,Brandenburg:2021bfx}.
For simplicity, we defer the study of the nonlinear effect produced by hydromagnetic turbulence
from cosmological phase transitions to future work, since it requires
a fully relativistic framework to reach plasma and/or Alfv\'en velocities
near speed of light, as expected for very strong sources.
There have been recent analytical works on nonhelical
and helical magnetogeneses during the reheating era
\cite{Sharma:2017eps,Sharma:2018kgs}, which circumvent
known difficulties such as the
backreaction and strong coupling problems \cite{Demozzi:2009fu}.
Numerical simulations of GWs from these magnetogeneses have also
been performed \cite{Brandenburg:2021pdv,Brandenburg:2021bfx} using the {\sc Pencil Code} \cite{pencil}.

This paper is organized as follows. In 
Sec.~\ref{sec:reheating_magnetogenesis}, we introduce the 
magnetogenesis models that will be used to generate the EM sources of GWs.
In Sec.~\ref{sec:GW_equations}, we discuss the relevant GW equation at the leading order beyond the standard linear case.
In Sec.~\ref{sec:numerical_simulations}, we present the simulation
parameters, discuss the results, and study the potential detectability
of the sources considered.
Finally, we conclude in Sec.~\ref{sec:conclusions}. 

We use the $(-\!\!+\!\!++)$ metric signature, set $c = 1$,
and normalize the critical energy density by the end of reheating 
$\eta_*$ to be unity, i.e., $\rho_\crit(\eta_*)= 1$, with
$\eta$ denoting conformal time.
We also adopt the convention that, for a function $F$, the Fourier transformation and its inverse read 
$\tilde{F}(\eta,\kk)=\int F(\eta,\xx)e^{-i\kk\cdot\xx}\dd^3\xx$ 
and $F(\eta,\xx)=\int \tilde F(\eta,\kk)e^{i\kk\cdot\xx}\dd^3\xx/(2\pi)^3$, respectively.
For long expressions in Fourier space, we also adopt the notation
${\cal F}(F(\eta, \xx)) = \tilde F (\eta, \kk)$.
In addition, Greek letters are employed for space-time indices running from 0 
to 3 and
Latin letters are used for space indices 1 to 3. 
Finally, variables by default refer to linear solutions but the ones with `nlin' superscript denote nonlinear solutions.

\section{Reheating magnetogenesis}
\label{sec:reheating_magnetogenesis}

A promising mechanism that can produce helical and nonhelical relic magnetic fields today is reheating magnetogenesis.
It is made possible by the coupling of the
inflaton field and the Maxwell Lagrangian
$\LLL^\EM = F_\munu F^\munu$ \cite{Ratra:1991bn}, where
$F_\munu\equiv\p_\mu A_\nu - \p_\nu A_\mu$ is the
Faraday tensor in terms of the EM four-potential $A_\mu$.
However, due to the conformal invariance of $\LLL^\EM$,
the EM amplitudes decay rapidly as the square of the scale factor during inflation. 
Therefore, violation of conformal invariance is required to obtain a growing EM field \cite{Turner:1987bw,Dolgov:1993vg}. 
This can be done, in the nonhelical case, via a coupling term of the form
$\LLL^\nhel = f^2F_\munu F^\munu$, with $f$ being a time-dependent function
that scales as
\begin{equation}
f(a)
\begin{cases}
\propto a^{\alpha}\;\;&(\text{during inflation}),\\
\propto a^{-\beta}\;\;&(\text{during reheating}),\\
= 1\;\;&(\text{during RD}),
\end{cases}
\label{eqn:f(a)}
\end{equation}
where
RD refers to the radiation-dominated era, which starts at
the end of reheating, 
$a$ is the scale factor, $\alpha$ is chosen to be 2, to avoid
the backreaction problem \cite{Sharma:2017eps}, or 1, which enables reheating temperatures above the electroweak (EW) scale \cite{Brandenburg:2021bfx} (see Sec.~\ref{sec:numerical_simulations}),
and $\beta > 0$ parameterizes the reheating temperature $\Tr$.
The detailed relations between $\Tr$ and $\beta$ can be found in Ref.~\cite{Sharma:2017eps,Sharma:2018kgs}.
Note that, in order to realize the standard electromagnetism, $f(a) = 1$ is
needed by the onset of radiation era, which we normalize to $\eta_* = 1$ \cite{Pol+18}.

On the other hand, helical magnetic fields can also be generated
by a modified coupling term of the form
$\LLL^\hel = \gam f^2 F_\munu\tilde{F}^\munu$, where $\gam\neq0$ is a
constant, $f$ obeys the same scaling relations as in
Eq.~(\ref{eqn:f(a)}),
and $\tilde{F}^\munu=\epsilon^{\mu\nu\alpha\beta}F_{\alpha\beta}/2$ is the dual of the Faraday tensor,
with $\epsilon^{\mu\nu\alpha\beta}$ being the fully antisymmetric tensor.

In addition, recall that the conformal time $\eta$ relates to the scale
factor $a$ as $a\propto \eta^{2/(1+3w)}$, where $w\equiv p/\rho$ is the
equation of state of the universe.
During the radiation era with $w = 1/3$, we have $a\propto\eta$.
During the matter-dominated reheating era, $w = 0$ gives
$a\propto\eta^2$.
Since we adopt the normalization such that $a_* = 1$ 
marks the onset of the radiation era at $\eta_* = 1$ \cite{Pol+18},
we have the following relations \cite{Sharma:2017eps}
\begin{equation}
a=
\begin{cases}
(\eta + 1)^2/4\;\;&(\text{during reheating}),\\
\eta\;\;&(\text{during RD}).
\end{cases}
\label{eqn:a(eta)}
\end{equation}

\subsection{Equations of motion for the four-potential}

The EM vector
potential $\tilde\AA$ can split into $\tilde A_+$ and $\tilde A_-$ polarization 
modes in Fourier space as \cite{Varshalovich:1988ye}
\begin{equation}
\tilde{\AA}(\eta,\kk) = \tilde{A}_+(\eta,\kk)\tilde{\ee}_+(\kk) + \tilde{A}_-(\eta,\kk)\tilde{\ee}_-(\kk),
\end{equation}
where
\begin{equation}
\tilde{\ee}_\pm(\kk) = - \frac{i}{\sqrt{2}} 
[\tilde{\ee}_1(\kk)\pm\tilde{\ee}_2(\kk)]
\label{eqn:pol_e}
\end{equation}
is the polarization basis satisfying $i\kk\times\tilde{\ee}_\pm
= \pm k\tilde{\ee}_\pm$. Here $k = |\kk|$ is the wave number, and
$\tilde{\ee}_1(\kk)$ and $\tilde{\ee}_2(\kk)$ are unit vectors
orthogonal to $\kk$ and to each other.

In Fourier space, the relevant equations for the four-potential
modes $\tilde A_\pm$ during reheating take the form \cite{Sharma:2018kgs,Brandenburg:2021bfx}
\begin{equation}
\tilde{\AAA}_\pm'' + \left(\kk^2\pm 2\gam k\frac{f'}{f} - \frac{f''}{f}\right)\tilde{\AAA}_\pm = 0,
\label{eqn:EOM_A}
\end{equation}
where $\tilde{\AAA}_\pm\equiv f\tilde{A}_\pm$ is the scaled four-potential and
$\tilde{\AAA}_\pm''\equiv\p^2\tilde{\AAA}_\pm/\p\eta^2$ is the second derivative with respect to conformal time $\eta$.
Here $\gam$ is a constant, and the terms involving the coupling function
$f$, scale factor $a$, and their derivatives can be expressed using
Eqs.~(\ref{eqn:f(a)}) and (\ref{eqn:a(eta)}), yielding
\begin{equation}
\frac{a''}{a} = \frac{2}{(\eta + 1)^2}\;,\;
\frac{f'}{f} = -\frac{2\beta}{\eta + 1}\;,\;
\frac{f''}{f} = \frac{2\beta(2\beta + 1)}{(\eta + 1)^2}.
\label{eqn:coupling_terms}
\end{equation}
In the helical basis, helicity is proportional to the difference between the
$+$ and $-$ modes.
This means that $\tilde\AAA_+ = \tilde\AAA_-$
in the nonhelical case, where $\gam=0$ and the modes are governed by $f''/f$ only.
In the helical case, $\gam=1$ and both $f'/f$ and $f''/f$ play a role in the solution.
The helical term in Eq.~(\ref{eqn:EOM_A}), proportional to $\gam$, leads to a difference in
the growth rates between $\tilde{\AAA}_+$ and $\tilde{\AAA}_-$. In our simulations, since there is enough time for the field to grow over many 
orders of magnitude, we consider only the dominant polarization mode in practice.

\section{GW Equations}
\label{sec:GW_equations}

\subsection{Leading-order nonlinear equation}
\label{ssec:leading-order_nonlinear_equation}

Given the metric $g_\munu = \bar g_\munu + h_\munu$, where $\bar g_\munu$
is the background metric tensor and $|h_\munu|\ll1$ correspond to high-frequency (compared to the slowly varying background)
small perturbations,
then the Einstein field equations (EFEs) can be expanded to arbitrary
orders in $h_\munu$, following Refs.~\cite{Isaacson67a,Isaacson67b};
see Ch. 20.3 of Ref.~\cite{Misner:1973prb} and Ch. 1 of Ref.~\cite{Maggiore07} for textbook references.

In general, EFEs read $G_\munu = \kap T_\munu$.
Here $T_\munu$ is the stress-energy tensor of the source,
$G_\munu\equiv R_\munu - \frac{1}{2}g_\munu R$ is the Einstein
tensor in terms of the Ricci tensor $R_\munu$ and Ricci scalar $R$,
and $\kap\equiv8\pi\GN$, where $\GN$ is Newton's gravitational constant.
As we are interested in nonlinear GWs, we expand $G_\munu$
up to, at least, quadratic order in $h_\munu$, giving
\begin{equation}
G_\munu = \bar G_\munu + G_\munu^{(1)} + G_\munu^{(2)} +
{\cal O} (h^3) = \kap T_\munu,
\end{equation}
where the superscripts (1) and (2) denote $\OOO(h)$ and $\OOO(h^2)$, 
respectively, and $\bar G_\munu$ depends on $\bar g_\munu$.
The equation for the $\OOO(h)$ terms can be recast as
\begin{gather}
G_\munu^{(1)} = \kap(T_\munu + t_\munu),
\label{eqn:einstein_tensor_1}
\end{gather}
where
we have introduced the effective stress of GWs, generally
referred to as the Landau-Lifshitz (LL) pseudotensor \cite{LanLif:1962},
$t_\munu=-G_\munu^{(2)}/\kappa$, with $G_\munu^{(2)}$ satisfying
\begin{equation}
G_\munu^{(2)} 
= R_\munu^{(2)} - \frac{1}{2}\bar g_\munu R^{(2)}.
\end{equation}

The RHS of Eq.~(\ref{eqn:einstein_tensor_1}) shows the two terms
sourcing GWs.
Rewriting with the trace-reversed perturbations
$\bar h_\munu\equiv h_\munu - \frac{1}{2}\bar g_\munu h$,
with $h = \bar g^\munu h_\munu$, and applying
the harmonic gauge $\p^\mu\bar h_\munu = 0$, the linearized Einstein
tensor becomes $G_\munu^{(1)} = -\frac{1}{2}\Box\bar h_\munu$,
where $\Box\equiv\p_\rho\p^\rho = -\p_t^2 + \DDel$ is the d'Alembert
operator.
Therefore, the linearized EFEs, i.e., omitting terms of order ${\cal O} (h^2)$ and higher, correspond to the usual GW equation:
\begin{equation}
    \Box \bar h_\munu = -2\kap T_\munu,
\end{equation}
while the expansion up to second order includes
the additional leading-order nonlinear term $t_\munu$.
Since we are interested in the resulting propagation of GWs
at the location of the observer, we can assume a flat background space-time
away from the source.

The second-order Ricci tensor is \cite{Maggiore07}
\begin{eqnarray}
R_\munu^{(2)} & = & \half \big(\half \p_\mu h_{\alpha\beta}\p_\nu h^{\alpha\beta} 
+ h^{\alpha\beta} \p_\mu \p_\nu h_{\alpha\beta} - h^{\alpha\beta}
\p_\nu \p_\beta h_{\alpha \mu} \nonumber \\
& - &  h^{\alpha\beta}\p_\mu \p_\beta h_{\alpha \nu} + 
h^{\alpha \beta}\p_\alpha \p_\beta h_\munu + \p^\beta h^\alpha_{\ \nu} \p_\beta h_{\alpha \mu} \nonumber \\
& - & \p^\beta h^\alpha_{\ \nu} \p_\alpha h_{\beta \mu} - 
\p_\beta h^{\alpha \beta} \p_\nu h_{\alpha \mu} +
\p_\beta h^{\alpha \beta} \p_\alpha h_\munu \nonumber  \\
& - & \p_\beta h^{\alpha \beta} \p_\mu h_{\alpha\nu} - 
\half \p^\alpha h\, \p_\alpha h_\munu  + 
\half \p^\alpha h\, \p_\nu h_{\alpha \mu}
\nonumber \\
& + & \half \p^\alpha 
h \p_\mu h_{\alpha \nu} \bigr).
\label{Rmunu2}
\end{eqnarray}
We choose the traceless and harmonic gauge, such that
$h = 0$ (and hence, $\bar h_\munu = 
h_\munu$, so we can drop the overbar from now on) and $\partial^\mu
h_\munu = 0$.

The energy density carried by GWs corresponds to the curvature they induce
on the background so we want to identify the terms
of the LL pseudotensor that contribute to the background metric
by averaging over the high frequencies that characterize the
strain perturbations $h_\munu$ compared to the background
$\bar g_\munu$.
Therefore, Eq.~(\ref{Rmunu2}) gets simplified to
(see Ref.~\cite{Maggiore07} for details)
\begin{equation}
    \bra{t_\munu} = - \frac{1}{\kappa} \bra{R_\munu^{(2)}} =  \frac{1}{4\kappa} \bra{\p_\mu h_{\alpha \beta}
    \p_\nu h^{\alpha \beta}},
\end{equation}
where the angle brackets refer to the aforementioned average.
If we require $k^i \tilde h_i = 0$,
then the strains are expressed in the traceless-transverse (TT) gauge,
such that the
gauge-invariant 00-component corresponds to the GW energy 
density \cite{Arnowitt:1961zza},
\begin{equation}
\bra{t_{00}} = \frac{1}{4\kappa} \bra{\p_t h_{ij}^{\rm TT}
    \p_t h_{ij}^{\rm TT}},
\end{equation}
and the spatial components are
\begin{equation}
\bra{t_{ij}} = \frac{1}{4\kappa} \bra{\p_i h_{lm}^{\rm TT}
    \p_j h_{lm}^{\rm TT}}.
\end{equation}

In the present context of continuous sources during cosmological epochs,
we consider the leading-order nonlinear GW equation, given in 
Eq.~(\ref{eqn:einstein_tensor_1}), and choose
the term that survives the average over high frequencies from all the nonlinear
terms in Eq.~\eqref{Rmunu2}, which corresponds to the memory effect
\cite{Will:1996zj,Favata09,Favata+10},
\begin{equation}
\Box h_{ij}^{\rm nlin} = -2\kap(T_{ij} + t_{ij}),
\label{GW_tLL}
\end{equation}
where the ``nlin'' superscript refers to the GW strains sourced by the
nonlinear terms up to leading-order and
\begin{equation}
t_{ij} = \frac{1}{4\kappa} \p_i h_{lm}^{\rm TT}
\p_j h_{lm}^{\rm TT}.
\label{t_LL}
\end{equation}
This is the equation, adapted to an expanding universe, that we use in our study of
the early universe in Sec.~\ref{ssec:EMGW_early_universe}.
We continue using the term ``memory effect,'' even though there may not be
any obvious permanent displacement of freely-falling test masses due to the GWB.

\subsection{Polarization modes of nonlinear solution}

We express the linear polarization modes of
the GW strain in Fourier space as
$\tilde{h}_{+, \times}(t, \kk) = \tilde e_{+, \times}^{\,ij} (\kk)
\tilde{h}_{ij}(t,\kk)$ \cite{Varshalovich:1988ye}.
The $+$ and $\times$ modes are obtained as a result of the TT projection,
and $\tilde e_{+, \times}^{\, ij}$ are the basis tensors,
\begin{equation}
    \tilde e_+^{\, ij} = \tilde e_i^1 \tilde e_j^1 -
    \tilde e_i^2 \tilde e_j^2, \quad
    \tilde e_\times^{\, ij} = \tilde e_i^1 \tilde e_j^2 +
    \tilde e_i^2 \tilde e_j^1,
\end{equation}
related to the helical polarization basis, given in Eq.~\ref{eqn:pol_e}
for a vector field, as
$\tilde e^{\, ij}_\pm = (\tilde e_+^{\, ij} \pm i \tilde e^{\, ij}_\times)/\sqrt{2}$ \cite{Caprini:2003vc}.
In the same way, we define $\tilde T_{+, \times}$ and $\tilde t_{+, \times}$ that source
the corresponding GW mode.

To study the polarization modes of $t_{ij}$, we take the WKB or eikonal approximation, i.e.,
we approximate the solution $\tilde h_\munu = {\cal C}_\munu
e^{i k_\alpha x^\alpha} \approx {\cal C}_\munu e^{i\phi}$,
being ${\cal C}_\munu$ the amplitudes of the perturbations,
and $\phi$ the constant GW phase.
The effective stress tensor of GWs, given in \Eq{t_LL},
can be expressed as [see Eq.~(4.5) of Ref.~\cite{Isaacson67b}]
\begin{equation}
    t_\munu^\WKB = \frac{{\cal C}^2}{4\kap}
    k_\mu k_\nu \sin^2 \phi,
\end{equation}
where ${\cal C}^2 = {\cal C}^\munu {\cal C}_\munu$ and $k^\mu = (\omega = k, k^i)$,
with $k^i$ being the characteristic wave number and $\omega$
the angular frequency,
or, equivalently, as
\begin{equation}
    t_{ij}^\WKB = t_{00} \hat k_i \hat k_j =
    \frac{1}{4\kap}  \bigl(\p_t h_{ab}^{\rm TT}
    \p_t h_{ab}^{\rm TT}\bigr) \hat k_i \hat k_j.
    \label{WKB}
\end{equation}
This expression is usually used in analytical studies of the
memory effect; see, e.g., Refs.~\cite{Favata:2008yd,Jenkins+21}.
Note that in our present work, we source the GWs
directly from $t_\munu$,
such that the WKB approximation is not necessary.
However, \Eq{WKB} is useful since it allows us to show
that only one polarization mode is sourced by the
effective stress tensor.
To see this, we Fourier transform \Eq{WKB} to get
\begin{equation}
    \tilde t_{ij}^\WKB = \frac{1}{4\kappa}
    {\cal F} \bigl(\p_t h_{ab}^{\rm TT}
    \p_t h_{ab}^{\rm TT}\bigr) * {\cal F}
    (\hat k_i \hat k_j),
\end{equation}
where $\cal F$ refers to the Fourier transform and is equivalent to
previous notation, which uses a tilde.
One can show that ${\cal F} (\hat k_i \hat k_j) \propto \delta_{ij}$ and, hence,
$\tilde t_{ij}^\WKB$ only has one non-zero mode,
since $\delta^+ = 1$ and $\delta^\times = 0$.
Hence, the GW \Eq{GW_tLL} can be expressed as
\begin{equation}
\Box \tilde h^{\rm nlin}_+ = -2\kap(\tilde T_+ + \tilde t_+), \quad 
 \Box \tilde h^{\rm nlin}_\times = -2\kap \tilde T_\times,
\label{hnlin}
\end{equation}
and the polarization modes of the nonlinear strains are $\tilde h^{\rm nlin}_+ =
\tilde h_+ + \Delta \tilde h_+$ and $\tilde h^{\rm nlin}_\times =
\tilde h_\times$, where $\tilde h_{+, \times}$ are the linear solutions sourced by
$\tilde T_{+, \times}$ and $\Delta \tilde h_+$ is the additional strain sourced by
$\tilde t_+$.

\subsection{EM-sourced GWs in the early universe}
\label{ssec:EMGW_early_universe}

For the early universe, we adopt conformal time $\eta$ such that $a \dd \eta =\dd t$
and set $\eta_* = H_*^{-1}$ as the end of reheating time,
where $H_*=\sqrt{\kap \rho_\crit(\eta_*)/3}$ is the Hubble
parameter at time $\eta_*$ with critical energy density $\rho_\crit(\eta_*)$. 

We normalize the conformal time $\bar\eta=\eta/\eta_*$,
such that $-1<\bar\eta<1$ corresponds to the reheating era, and
we use comoving and normalized wave vector $\bar\kk=\kk/(a H_*)$.
Since we study the GW relics at the end of reheating, i.e., at the onset of radiation-dominated era, we approximate $\rho_\crit$ by the radiation
energy density as $\rho_\crit(\eta_*)\simeq\EEE_\rad(\eta_*)=\pi^2\gr(T_*)\kB^4T_*^4/(30\hbar^3)$, 
where $\gr(T_*)$ is the number of relativistic degrees of freedom at temperature 
$T_*$, $\kB$ is the Boltzmann constant, and $\hbar$ is the reduced Planck constant.
We also scale the strain $\bar h_{ij}=a h_{ij}$ and normalize the comoving stress of the source
$\bar T_{ij} = a^4 T_{ij}/\rho_\crit$.
Then dropping all the overbars, the resulting GW equation \cite{Pol+18},
with the leading-order nonlinear term added, reads\footnote{
In Eq.~\eqref{hnlin}, we set $\tilde t_\times=0$ based on the eikonal approximation.
However, we compute $\tilde t_\times$ in the numerical simulations, so we
keep this term in \Eq{eqn:GW_scaled1}.}
\begin{equation}
\Bigl(\tilde{h}^{\rm nlin}_{+, \times}\Bigr)'' + \biggl(k^2 - \frac{a''}{a}\biggr)\tilde{h}_{+, \times}^{\rm nlin} =
\frac{6}{a}\bigl(\tilde{T}_{+, \times} +  \tilde{t}_{+, \times} \bigr),
\label{eqn:GW_scaled1}
\end{equation}
where $a''/a$ is given in Eq.~(\ref{eqn:coupling_terms}).
See Refs.~\cite{Maggiore07,Caprini+18} for details and Ref.~\cite{Pol+18} for
the implementation in the {\sc Pencil Code} \cite{pencil}.
The EM stress components in physical space are assembled as
\begin{equation}
T_{ij} = f^2(B_iB_j + E_iE_j),
\label{Tij}
\end{equation}
where $B_i$ and $E_i$ are components of the magnetic field $\BB=\Del\times\AA$ and electric field $\EE=-\p\AA/\p\eta$.

In connection with potential observations, we obtain the present day values of
the GW energy and helicity spectrum in the forms $h_0^2\OmGW$ and $h_0^2\XiGW$, 
which are independent of the uncertainty in the present day Hubble parameter $H_0 = 100 \, h_0 \km \s^{-1} \Mpc^{-1}$,
with $h_0\approx0.7$ \cite{Planck:2018vyg},
\begin{eqnarray}
\OmGW(f) & = &
\biggl(\frac{H_*}{H_0}\biggr)^2\biggl(\frac{a_*}{a_0}\biggr)^4k\EGW(k),
\label{eqn:OmGW_f}\\
\XiGW(f) & = &
\biggl(\frac{H_*}{H_0}\biggr)^2\biggl(\frac{a_*}{a_0}\biggr)^4k\HGW(k),
\label{eqn:XiGW_f}
\end{eqnarray}
where $H_*$ and $a_*$ are the Hubble parameter and scale factor at the end of 
reheating, and $H_0$ and $a_0$ are their present day values.
Here $f=a_* H_*k/(2\pi a_0)$ is the physical frequency today.
The GW spectra $\EGW (k)$ and $\HGW (k)$ are computed from the time derivatives
of the strains as
\begin{align}
    \EGW (k) & \, = \int\Bigl(\bigl|\dot{\tilde{h}}_+\bigr|^2+\bigl|\dot{\tilde{h}}_\times
    \bigr|^2\Bigr)k^2\dd\Omega_k,
    \\
    \HGW (k) & \, = \int2\,\Im\,\Bigl(\dot{\tilde{h}}_+\dot{\tilde{h}}_\times^*
    \Bigr)k^2\dd\Omega_k,
\end{align}
where $\Omega_k$ is the solid angle of the shell with size $k$.

\subsection{Order-of-magnitude estimates}
\label{ssec:OME}

\subsubsection{Effects on the Maxwell equation}
\label{sssec:Amunu}

Nonlinear effects at the order $\OOO(h^2)$ can potentially enter GW production at the order $\OOO(h)$ in two ways.
The first way, as has been shown in Sec.~\ref{ssec:leading-order_nonlinear_equation}, is to have a
term $t_{ij}=\OOO(h^2)$, given in \Eq{t_LL}, directly on the RHS of the
standard GW equation, i.e., \Eq{GW_tLL},
thereby acting as an additional source.
The second way is to
see how $\OOO(h^2)$ changes the regular EM source and whether in turn
the modified EM source could also affect GW productions at the linear
order $\OOO(h)$.

For the latter, we use the first of Maxwell's equations in curved spacetime
\begin{equation}
\del_\mu F^\munu = \frac{1}{\sqrt{-g}}\p_\mu\bigl(\sqrt{-g}F^\munu \bigr)
= \mu_0 j^\nu = 0.
\label{eqn:Maxwell}
\end{equation}
We have set the current to vanish during the reheating era, where the relevant Maxwell's equations are in vacuum.
This can be rewritten as
\begin{eqnarray}
\p_\mu F^\munu & = & -\dfrac{\p_\mu\sqrt{-g}}{\sqrt{-g}}F^\munu\nonumber\\
& = & -\biggl(\dfrac{1}{2}\p_\mu(\bar g^{\rho\sig}h_{\rho\sig}) + \OOO(h^2)\biggr) F^\munu,\\
\Box A^\nu & = & \Big(\OOO(h) + \OOO(h^2)\Big)\OOO(A),
\label{eqn:Maxwell_dim}
\end{eqnarray}
where we have used
\begin{equation}
\frac{\p_\mu\sqrt{-g}}{\sqrt{-g}} 
= \frac{1}{2}\bar g^{\rho\sig}\p_\mu g_{\rho\sig}
= \frac{1}{2}\p_\mu(\bar g^{\rho\sig}h_{\rho\sig}) + \OOO(h^2),
\end{equation}
and applied the Lorenz gauge condition $\p_\mu A^\mu = 0$,
such that $\p_\mu F^\munu = \Box A^\nu$.

Therefore, Eq.~(\ref{eqn:Maxwell_dim}) shows that the leading-order nonlinearity in the GWs,
of the order $\OOO(h^2)$, does modify the four-potential, which registers a difference
\begin{equation}
\Delta A = \OOO(h^2)\OOO(A).
\end{equation}
However, the result of such additional EM amplitude $\EEEM$, which is proportional to $A^2$; see \Eq{Tij}, introduces a modification to the production
of GWs at $h = {\cal O} (\EEEM) = \OOO(A^2)$; see \Eq{eqn:GW_scaled1}, of larger order than quadratic in $h$.
For this reason, we only consider the additional GWs produced by their
self-backreaction at $t_{ij}=\OOO(h^2)$, rather than the backreaction from the vector
potential at $\OOO((\Delta A)^2) = \OOO(h^4)
\OOO(A^2) = \OOO(h^5)$.

\subsubsection{GW energy}
\label{sssec:EEGW}

The standard GW equation is linear in the strain and its sourcing energy, 
i.e., $h = \OOO(\EEEM)$.
Since GW energy is quadratic in strain, i.e., $\EEGW = \OOO(h^2)$, we have a quadratic relation
\begin{equation}
\EEGW = \OOO(\EEEM^2),
\label{eqn:OME_EEGW}
\end{equation}
which has been verified by a number of earlier simulations \cite{RoperPol:2019wvy,Brandenburg:2021aln,Brandenburg:2021tmp,Pol:2021uol}.
Adding nonlinearities of the order $\OOO(h^2)$ to the source means that
$h^\nlin = h + \OOO(h^2)$. 
Therefore, the nonlinear GW energy relates to the source energies as
\begin{align}
\EEGW^\nlin
& = \bigl[(\EEGW)^{1/2} + \OOO(h^2)\bigr]^2  \nonumber \\
& = \EEGW + \EEGW^{1/2} \OOO(h^2) + {\cal O}(h^4) \nonumber \\
& = \EEGW + \OOO(\EEEM^3) + {\cal O}(\EEEM^4),
\label{oom_EEGW}
\end{align}
which implies that the leading-order energy difference
$\Delta\EEGW\equiv\EEGW^\nlin-\EEGW$ should be proportional to the
cube of the sourcing energy, i.e.,
\begin{equation}
\Delta\EEGW = \OOO(\EEEM^3).
\label{eqn:OME_dEEGW}
\end{equation}
This relation will be verified and quantified numerically in Sec.~\ref{ssec:dependence_on_EEEM}.

\subsubsection{GW polarization}
\label{sssec:PPGW}

The GW polarization is defined to be \cite{Kahniashvili:2005qi}
\begin{equation} 
\PPP_\GW(k) = \frac{\HGW (k)}{\EGW (k)}.
\label{eqn:GWpol}
\end{equation}
The differences in GW spectra due to the nonlinear solution; see \Eq{hnlin}, are
\begin{align}
\EGW^{\rm nlin} (k) - \EGW (k) \propto & \,  \Delta \tilde h_+ \tilde h_+^* + \tilde h_+ \Delta \tilde h_+^*
\nonumber \\
= & \, {\cal O} (h^3) = {\cal O} (\EEEM^3), \label{EGWnlin_EGW} \\
\HGW^{\rm nlin} (k) - \HGW (k) \propto & \, \Delta \tilde h_+ \tilde h_\times^* -
\Delta \tilde h_+^* \tilde h_\times \nonumber \\
= & \, {\cal O} (h^3) = {\cal O} (\EEEM^3).
\label{HGWnlin_HGW}
\end{align}
Note that \Eq{EGWnlin_EGW} is consistent with \Eq{oom_EEGW}.
We can now express the relative differences in the GW polarization due to the addition
of the nonlinear term (omitting the $k$ dependence to simplify the notation) as
\begin{align}
\frac{\PPGW -  \PPGW^{\rm nlin}}{\PPGW} =  & \, 1 - \frac{\HGW^{\rm nlin} \EGW}
{ \HGW \EGW^{\rm nlin}} \nonumber \\
= & \, 1 - \frac{1 + \HGW^{-1}\, {\cal O} (h^3)}
{1 + \EGW^{-1}\, {\cal O} (h^3)} \nonumber \\
= & \, 1 - (1 + \HGW^{-1}\, {\cal O} (h^3))(1 - \EGW^{-1} \,{\cal O} (h^3)) \nonumber \\
= & \,  \EGW^{-1} \, {\cal O} (h^3)  - \HGW^{-1}\, {\cal O} (h^3)  \nonumber \\ =
& \, {\cal O} (h) - \PPGW^{-1} {\cal O} (h) \nonumber \\ = 
& \, {\cal O} (h) = {\cal O} (\EEEM).
\label{PGW_order}
\end{align} 
We then expect a decrease in polarization due to the
nonlinear contributions proportional to the EM
energy density and to the polarization obtained in
the linearized approach, which will be verified in
Sec.~\ref{ssec:dependence_on_beta}.
We find that the sign of the difference and its exact value
depend on the balance between the two terms of order ${\cal O} (h)$,
which correspond to those given in \Eqs{EGWnlin_EGW}{HGWnlin_HGW}
divided by $\HGW (k)$ and $\EGW (k)$, respectively.

\section{Numerical simulations}
\label{sec:numerical_simulations}

Tables~\ref{tab:scaling_param} and \ref{tab:runs} summarize the relevant simulation parameters and their final output.
In this work, all simulations are performed on $n^3=512^3$ mesh points with the smallest wave number $k_1 = 1$, 
except in series E, where $k_1=0.2$.
The EM energy spectra peak at wave numbers $k_*$
within the simulation domain,
i.e., $k_1<k_*<\kNy$, where $\kNy=nk_1/2$ is the Nyquist wave number.
We conduct five series of simulations: two nonhelical series (A and B) and three helical series (C, D, and E).

\begin{table}[b]
\caption{Choice of model parameters, initial magnetic spectrum for $k\leq k_*$,
and energy dilution factors.}
\centering
\renewcommand{\arraystretch}{1.25}
\begin{tabular}{c|cccc|c|c}
Series & $\Tr[\GeV]$ & $\gam$ & $\beta$ & $\gr$ & $E_{\rm M}(k)$ & $(H_*/H_0)^2(a_*/a_0)^4$ \\\hline
A & 100 & 0 & 7.3 & 106 & $\propto k^3$ & $1.6\times10^{-5}$\\
B & 0.15 & 0 & 2.7 & 15 & $\propto k^3$ & $3.1\times10^{-5}$\\
\hline
C & 8 & 1 & 7.3 & 86 & $\propto k^3$ & $1.7\times10^{-5}$\\
D & 0.12 & 1 & 2.7 & 20 & $\propto k^3$ & $2.8\times10^{-5}$\\
E & $3\times10^5$ & 1 & 1.7 & 106 & $\propto k^5$ & $1.6\times10^{-5}$\\
\end{tabular}
\label{tab:scaling_param}
\end{table}

\begin{table*}
\caption{Summary of simulations and relevant quantities.
A', B', C', D', and E' runs adopt the same parameters as A, B, C, D, 
and E, respectively, but are run up to $\eta=10$ and we show their
saturated values.
For the rest of the runs, the values correspond to $\eta=1$.
In all cases, $k_*$ is shown at $\eta = 1$.}
\centering
\renewcommand{\arraystretch}{1.2}
\begin{tabular}{c|l|c|c|lll|rrr}
Runs &  \hspace{3.5mm} $B_0$  &
$\EEEM$ & $k_*(1)$ & \hspace{2mm} $\EEGW$ &
$\hspace{2mm} \Delta\EEGW$ & \hspace{-2.5mm} $\Delta\EEGW/\EEGW^\nlin$
& \hspace{2.5mm} $\PPGW^\nlin$ &  $\Delta\PPGW$ \hspace{2.5mm} & $\Delta\PPGW/\PPGW$ \hspace{-2mm} \\\hline
A1 & $3.3 \times 10^{-19}$ & 0.02 & $7.5$ & $1.2 \times 10^{-5}$ &
$2.7 \times 10^{-10}$ & $2.3 \times 10^{-5}$ &
$-0.108$ & $1.8 \times 10^{-7}$ & $-1.6 \times 10^{-6}$ \\
\blue{\,A1'} & \blue{$3.3 \times 10^{-19}$} & \blue{0.02} & \blue{$7.5$} & \blue{$7.7 \times 10^{-6}$} &
\blue{$1.6 \times 10^{-9}$} & \blue{$2.1 \times 10^{-4}$} &
\blue{$-0.108$} & \blue{$2.5 \times 10^{-4}$} & \blue{$-2.3 \times 10^{-3}$} \\
A2 & $7.5 \times 10^{-19}$ & 0.1 & $7.5$ & $3.2 \times 10^{-4}$ &
$3.8 \times 10^{-8}$ & $1.2 \times 10^{-4}$ &
$-0.108$ & $9.2 \times 10^{-7}$ & $-8.6 \times 10^{-6}$ \\
\blue{\,A2'} & \blue{$7.5 \times 10^{-19}$} & \blue{0.1} & \blue{$7.5$} & \blue{$2.1 \times 10^{-4}$} &
\blue{$2.7 \times 10^{-7}$} & \blue{$1.3 \times 10^{-3}$} &
\blue{$-0.109$} & \blue{$1.3 \times 10^{-3}$} & \blue{$-1.2 \times 10^{-2}$} \\
A3 & $2.4 \times 10^{-18}$ & 1 & $7.5$ & $3.4 \times 10^{-2}$ &
$4.0 \times 10^{-5}$ & $1.2 \times 10^{-3}$ &
$-0.108$ & $9.6 \times 10^{-6}$ & $-8.9 \times 10^{-5}$ \\
A4 & $7.5 \times 10^{-18}$ & 10 & $7.5$ & $3.2 \times 10^{0}$ &
$3.8 \times 10^{-2}$ & $1.2 \times 10^{-2}$ &
$-0.108$ & $1.0 \times 10^{-4}$ & $-9.3 \times 10^{-4}$ \\
B1 & $2.7 \times 10^{-7}$ & 0.02 & $2.9$ & $1.1 \times 10^{-4}$ &
$9.5 \times 10^{-8}$ & $8.7 \times 10^{-4}$ &
$-0.405$ & $6.2 \times 10^{-5}$ & $-1.5 \times 10^{-4}$ \\
\green{\,B1'} & \green{$2.7 \times 10^{-7}$} & \green{0.02} & \green{$2.9$} & \green{$7.9 \times 10^{-5}$} &
\green{$1.8 \times 10^{-7}$} & \green{$2.2 \times 10^{-3}$} &
\green{$-0.380$} & \green{$1.8 \times 10^{-3}$} & \green{$-4.8 \times 10^{-3}$} \\
B2 & $6.0 \times 10^{-7}$ & 0.1 & $2.9$ & $2.6 \times 10^{-3}$ &
$1.1 \times 10^{-5}$ & $4.3 \times 10^{-3}$ &
$-0.406$ & $3.0 \times 10^{-4}$ & $-7.5 \times 10^{-4}$ \\
\green{\,B2'} & \green{$6.0 \times 10^{-7}$} & \green{0.1} & \green{$2.9$} & \green{$1.9 \times 10^{-3}$} &
\green{$2.9 \times 10^{-5}$} & \green{$1.5 \times 10^{-2}$} &
\green{$-0.386$} & \green{$8.2 \times 10^{-3}$} & \green{$-2.2 \times 10^{-2}$} \\
B3 & $1.9 \times 10^{-6}$ & 1 & $2.9$ & $2.7 \times 10^{-1}$ &
$1.2 \times 10^{-2}$ & $4.4 \times 10^{-2}$ &
$-0.408$ & $2.8 \times 10^{-3}$ & $-6.9 \times 10^{-3}$ \\
B4 & $6.0 \times 10^{-6}$ & 10 & $2.9$ & $2.6 \times 10^{1}$ &
$1.9 \times 10^{1}$ & $4.2 \times 10^{-1}$ &
$-0.411$ & $5.5 \times 10^{-3}$ & $-1.4 \times 10^{-2}$ \\\hline
C1 & $1.7 \times 10^{-24}$ & 0.02 & $17.8$ & $2.4 \times 10^{-6}$ &
$7.2 \times 10^{-10}$ & $3.0 \times 10^{-4}$ &
$0.942$ & $3.9 \times 10^{-5}$ & $4.1 \times 10^{-5}$ \\
\orange{\,C1'} & \orange{$1.7 \times 10^{-24}$} & \orange{0.02} & \orange{$17.8$} & \orange{$3.7 \times 10^{-6}$} &
\orange{$1.1 \times 10^{-10}$} & \orange{$2.9 \times 10^{-5}$} &
\orange{$0.971$} & \orange{$-1.0 \times 10^{-6}$} & \orange{$-1.0 \times 10^{-6}$} \\
C2 & $3.9 \times 10^{-24}$ & 0.1 & $17.8$ & $6.7 \times 10^{-5}$ &
$1.0 \times 10^{-7}$ & $1.6 \times 10^{-3}$ &
$0.942$ & $2.0 \times 10^{-4}$ & $2.2 \times 10^{-4}$ \\
\orange{\,C2'} & \orange{$3.9 \times 10^{-24}$} & \orange{0.1} & \orange{$17.8$} & \orange{$1.0 \times 10^{-4}$} &
\orange{$1.7 \times 10^{-8}$} & \orange{$1.7 \times 10^{-4}$} &
\orange{$0.971$} & \orange{$3.0 \times 10^{-6}$} & \orange{$3.1 \times 10^{-6}$} \\
C3 & $1.2 \times 10^{-23}$ & 1 & $17.8$ & $6.0 \times 10^{-3}$ &
$9.0 \times 10^{-5}$ & $1.5 \times 10^{-2}$ &
$0.940$ & $2.0 \times 10^{-3}$ & $2.1 \times 10^{-3}$ \\
C4 & $3.9 \times 10^{-23}$ & 10 & $17.8$ & $6.7 \times 10^{-1}$ &
$1.2 \times 10^{-1}$ & $1.5 \times 10^{-1}$ &
$0.918$ & $2.4 \times 10^{-2}$ & $2.6 \times 10^{-2}$ \\
D1 & $2.4 \times 10^{-9}$ & 0.02 & $6.7$ & $1.6 \times 10^{-5}$ &
$4.1 \times 10^{-8}$ & $2.5 \times 10^{-3}$ &
$0.944$ & $2.8 \times 10^{-4}$ & $3.0 \times 10^{-4}$ \\
\red{\,D1'} & \red{$2.4 \times 10^{-9}$} & \red{0.02} & \red{$6.7$} & \red{$2.7 \times 10^{-5}$} &
\red{$1.6 \times 10^{-8}$} & \red{$6.0 \times 10^{-4}$} &
\red{$0.972$} & \red{$8.0 \times 10^{-5}$} & \red{$8.2 \times 10^{-5}$} \\
D2 & $5.4 \times 10^{-9}$ & 0.1 & $6.7$ & $4.2 \times 10^{-4}$ &
$5.4 \times 10^{-6}$ & $1.3 \times 10^{-2}$ &
$0.943$ & $1.5 \times 10^{-3}$ & $1.5 \times 10^{-3}$ \\
\red{\,D2'} & \red{$5.4 \times 10^{-9}$} & \red{0.1} & \red{$6.7$} & \red{$6.9 \times 10^{-4}$} &
\red{$2.3 \times 10^{-6}$} & \red{$3.4 \times 10^{-3}$} &
\red{$0.972$} & \red{$6.2 \times 10^{-4}$} & \red{$6.4 \times 10^{-4}$} \\
D3 & $1.7 \times 10^{-8}$ & 1 & $6.7$ & $4.1 \times 10^{-2}$ &
$5.8 \times 10^{-3}$ & $1.2 \times 10^{-1}$ &
$0.927$ & $1.7 \times 10^{-2}$ & $1.8 \times 10^{-2}$ \\
D4 & $5.4 \times 10^{-8}$ & 10 & $6.7$ & $4.2 \times 10^{0}$ &
$2.4 \times 10^{1}$ & $8.5 \times 10^{-1}$ &
$0.672$ & $2.7 \times 10^{-1}$ & $2.9 \times 10^{-1}$ \\
E1 & $4.5 \times 10^{-6}$ & 0.02 & $4.3$ & $5.3 \times 10^{-5}$ &
$3.4 \times 10^{-7}$ & $6.3 \times 10^{-3}$ &
$0.909$ & $8.1 \times 10^{-4}$ & $8.9 \times 10^{-4}$ \\
\purple{\,E1'} & \purple{$4.5 \times 10^{-6}$} & \purple{0.02} & \purple{$4.3$} & \purple{$9.4 \times 10^{-5}$} &
\purple{$3.4 \times 10^{-7}$} & \purple{$3.6 \times 10^{-3}$} &
\purple{$0.941$} & \purple{$4.6 \times 10^{-4}$} & \purple{$4.8 \times 10^{-4}$} \\
E2 & $1.0 \times 10^{-5}$ & 0.1 & $4.3$ & $1.3 \times 10^{-3}$ &
$4.2 \times 10^{-5}$ & $3.2 \times 10^{-2}$ &
$0.905$ & $4.4 \times 10^{-3}$ & $4.9 \times 10^{-3}$ \\
\purple{\,E2'} & \purple{$1.0 \times 10^{-5}$} & \purple{0.1} & \purple{$4.3$} & \purple{$2.3 \times 10^{-3}$} &
\purple{$4.5 \times 10^{-5}$} & \purple{$1.9 \times 10^{-2}$} &
\purple{$0.938$} & \purple{$3.3 \times 10^{-3}$} & \purple{$3.5 \times 10^{-3}$} \\
E3 & $3.2 \times 10^{-5}$ & 1 & $4.3$ & $1.4 \times 10^{-1}$ &
$7.3 \times 10^{-2}$ & $3.5 \times 10^{-1}$ &
$0.805$ & $1.0 \times 10^{-1}$ & $1.1 \times 10^{-1}$ \\
\end{tabular}
\label{tab:runs}
\end{table*}

In Table~\ref{tab:scaling_param}, we list the relevant model parameters for 
all series of 
the 
nonhelical ($\gamma = 0$) and helical ($\gamma = 1$) magnetogenesis scenarios
considered in the present work.
Note that, for the same values of $\beta$, despite the different corresponding temperatures $\Tr$, due to helicity, the energy dilution factors
$(H_*/H_0)^2(a_*/a_0)^4$ remain similar.
This is because the dilution factor is proportional to $\gr^{-1/3}$.

There are five variants for series A, B, C, and D each, 
corresponding to final EM energy $\EEEM\in\{0.02,0.1,1,10\}$, 
and four variants for series E, where $\EEEM=10$ is absent since the GW solution diverged in this case.
In reality, we expect the EM energy density to be $\EEEM \lesssim 0.1$,
which is generally believed to be the upper limit on the additional relativistic 
components based on the abundance of light elements imposed by the constraints of Big Bang
Nucleosynthesis (BBN) \cite{Shvartsman:1969mm,Grasso:1996kk}.
However, we also take unrealistically large values in the present work to verify the scaling of the memory effect.

The current study mainly focuses on the nonlinear features of GWs during the
reheating era, when the sourcing magnetogenesis takes place concurrently.
However, we are also interested in the resulting nonlinear solutions up to our present time to study their potential detectability.
For this reason, we choose additional runs per series with $\EEEM = 0.02$ and $0.1$, and evolve them beyond the end of reheating at $\eta=1$.
The simulation is then continued up to $\eta = 10$, but with the EM source being
turned off at $\eta=1$.
In these longer runs, we observe that the resulting GW energy density,
polarization, and spectra, become oscillatory around stationary solutions, just
as in previous numerical simulations \cite{RoperPol:2019wvy,Pol:2021uol}.
Hence, we average the results over the oscillations in time between
$\eta = 2$ and $10$ and assume the GW energy density to evolve
proportional to $a^{-4}$ up to our present time.
The new runs are denoted as A', B', C', D', and E'.
To avoid any additional discontinuities at $\eta=1$,
we continue using $a=(\eta+1)^2/4$, i.e., no attempt is
made to model the subsequent radiation-dominated era.
Note that $a''/a$ is 0.5 at $\eta = 1$ and decreases to zero as $\eta^{-2}$.
All these runs produce a certain change of the GW energy,
but it is only weakly related to the effect of $t_{ij}$.

For each variant, with fixed $\beta$ and $\EEEM$, we then obtain the total GW
energy density $\EEGW$ and polarization\footnote{
The total GW energy density and polarization are computed
from the GW spectra as $\EEGW = \int \EGW (k) \dd k$ and
$\PPGW = \int \HGW (k) \dd k/\int \EGW (k) \dd k$.} $\PPGW$
for both the linear and nonlinear cases.
Table~\ref{tab:runs} summarizes the relevant simulation parameters and their output at $\eta=1$,
where
the nonlinear effects are
represented by $\Delta\EEGW\equiv\EEGW^\nlin-\EEGW$ and
$\Delta\PPGW \equiv \PPGW-\PPGW^{\rm nlin}$.
For the continued runs, we show their values averaged between $\eta=2$ and 10.

In the following, we introduce the relevant model parameters and initial conditions in Sec.~\ref{sec:choice_of_param} and we
present the nonlinear 
effects to the GW and polarization spectra
for different initial fields $B_0$ 
and reheating parameters $\beta$ in Sec.~\ref{ssec:dependence_on_beta}.
Their relation with the sourcing energy $\EEEM$
is parameterized in Sec.~\ref{ssec:dependence_on_EEEM}, 
their long-term evolution is shown in Sec.~\ref{ssec:late_evol}, 
and the possibility for detection is considered in Sec.~\ref{ssec:observation}.
Only the results of Sec.~\ref{ssec:observation} correspond to observables at the
present time, while previous results are shown in the scaled and normalized
units presented in \Secs{sec:reheating_magnetogenesis}{sec:GW_equations}.

\subsection{Choice of parameters}
\label{sec:choice_of_param}

Here we briefly explain the choice of parameters used for this work.
The values of $\alpha$ and $\beta$, governing the evolution of the coupling function $f$; see \Eq{eqn:f(a)}, and the magnetic field initial conditions are all chosen in line with 
Refs.~\cite{Brandenburg:2021pdv,Brandenburg:2021bfx}.
For series A and B, we choose 
$\alpha = 2$ to avoid the backreaction problem \cite{Sharma:2017eps}, and $\beta\in\{7.3, 2.7\}$, corresponding to reheating
temperatures of $\Tr\in\{100, 0.15\} \GeV$, i.e., the energy levels of 
EW and QCD phase transitions.
For helical scenarios, series C and D carry the same value of $\alpha$ and
$\beta$ values are maintained for comparison with series A and B,
although now they correspond to $\Tr\in\{8,0.12\} \GeV$.
The helical series E takes $\alpha=1$ to enable a higher-than-EW
reheating temperature
and $\beta=1.7$, which corresponds to $\Tr=3\times10^5 \GeV$;
see Ref.~\cite{Brandenburg:2021bfx} for details.
For series A, B, C, and D, the initial conditions at $\eta = \eta_{\rm ini}=-0.99$ for
Eq.~\eqref{eqn:EOM_A}
are random, Gaussian-distributed, magnetic fields with strengths $B_0\ll1$
and energy spectra
\begin{equation}
E_{\rm M}(k)
\propto
\begin{cases}
k^3\;\;&(k\leq k_*(\eta_{\rm ini})),\\
k^{1-4\beta}\;\;&(k>k_*(\eta_{\rm ini})),
\end{cases}
\label{eqn:EM_ini}
\end{equation}
and
\begin{equation}
k_*(\eta)=\frac{2\beta}{\eta+1}\Bigg(\gam+\sqrt{1+\gam^2+\frac{1}{2\beta}}\Bigg)
\label{eqn:kstar}
\end{equation}
indicates the peak of the EM energy spectrum.
For series E, we initialize with $E_{\rm M}(k)\propto k^5$ for all wave numbers. 
However, \Eq{eqn:kstar} is still valid for runs in series E at the end of reheating
as they develop $k_*(\eta)$ during their evolution.

\subsection{Dependence on $B_0, \beta$, and helicity}
\label{ssec:dependence_on_beta}

\begin{figure*}
\includegraphics[width=.36\textwidth]{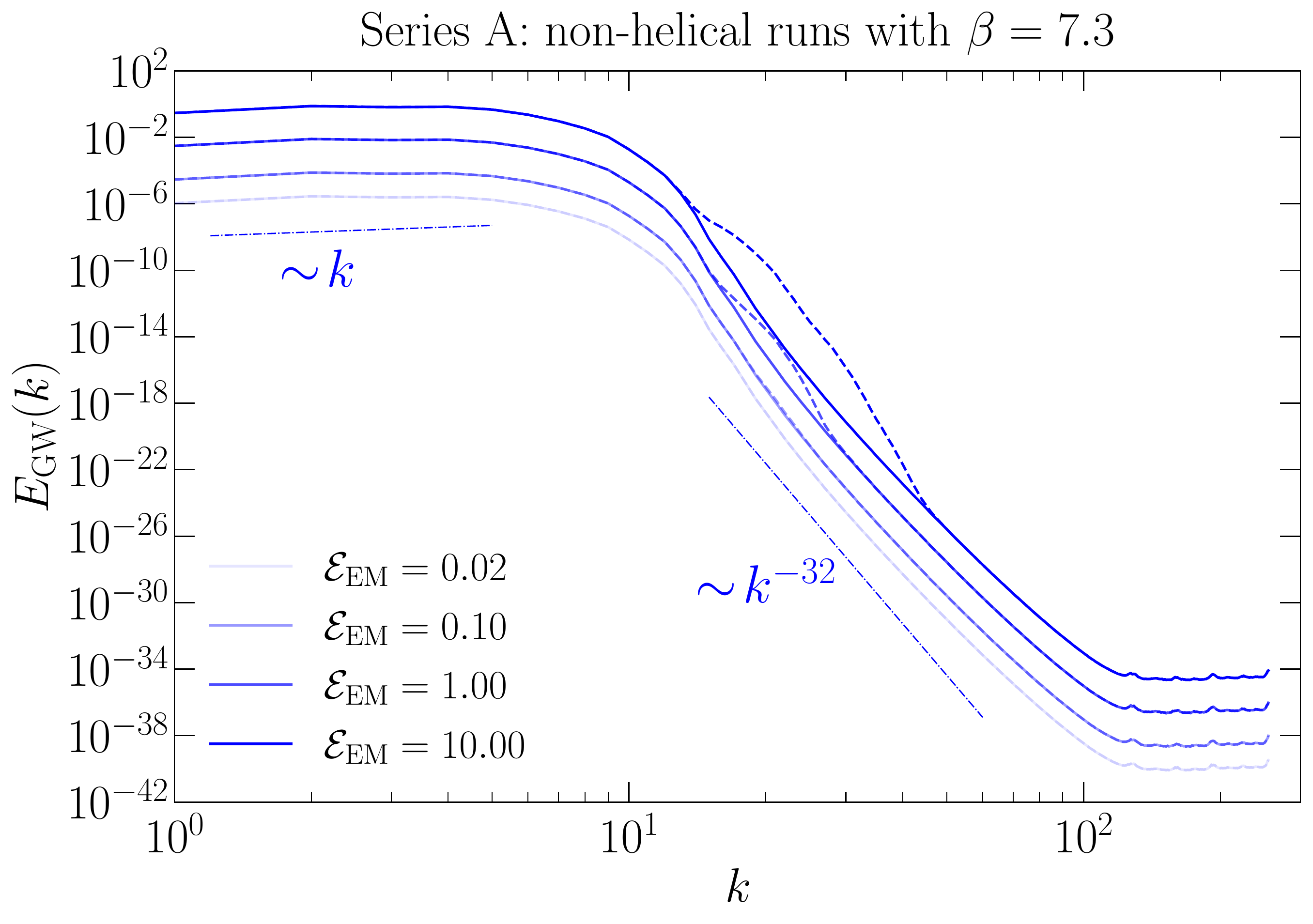}
\includegraphics[width=.40\textwidth]{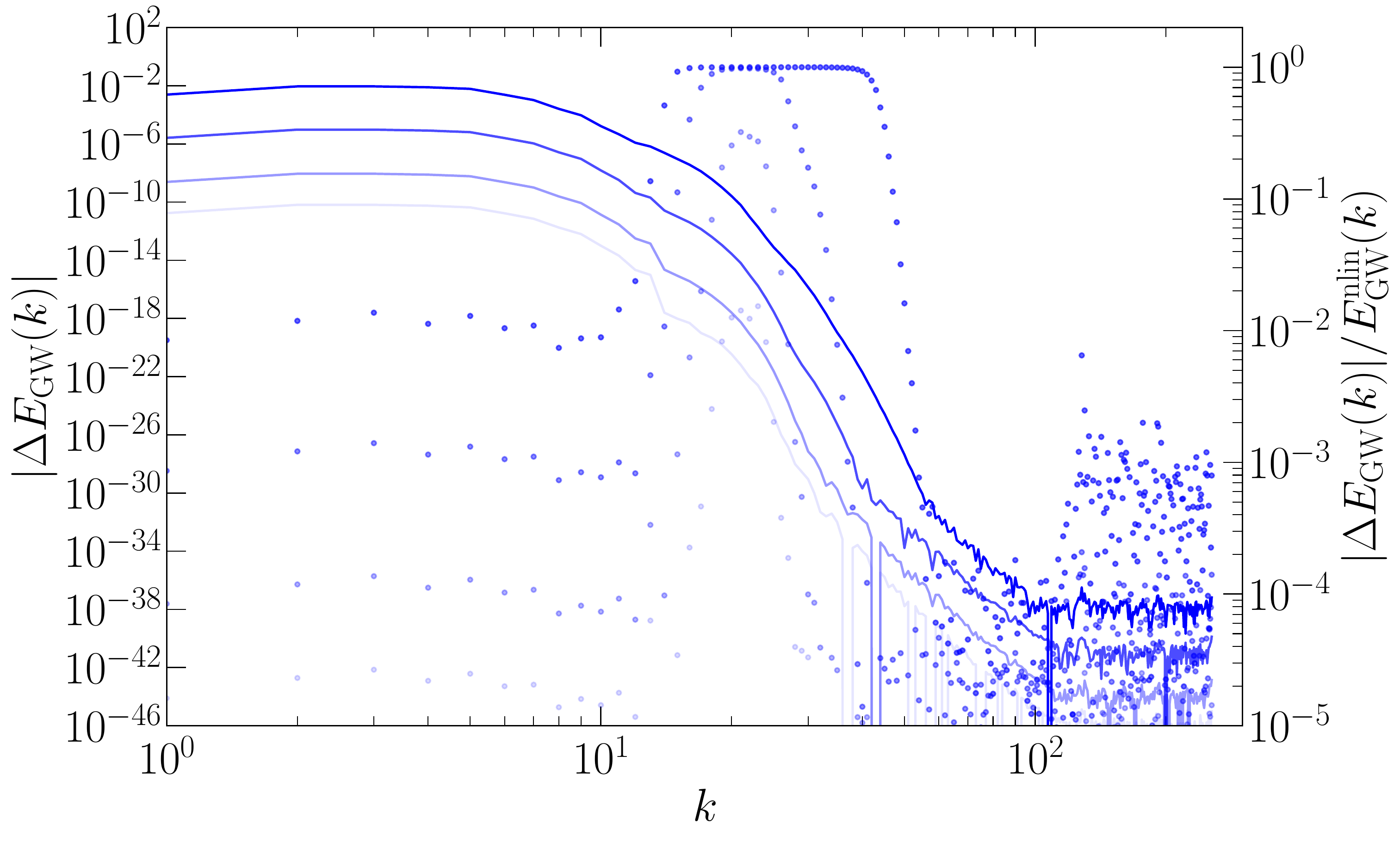}
\includegraphics[width=.36\textwidth]{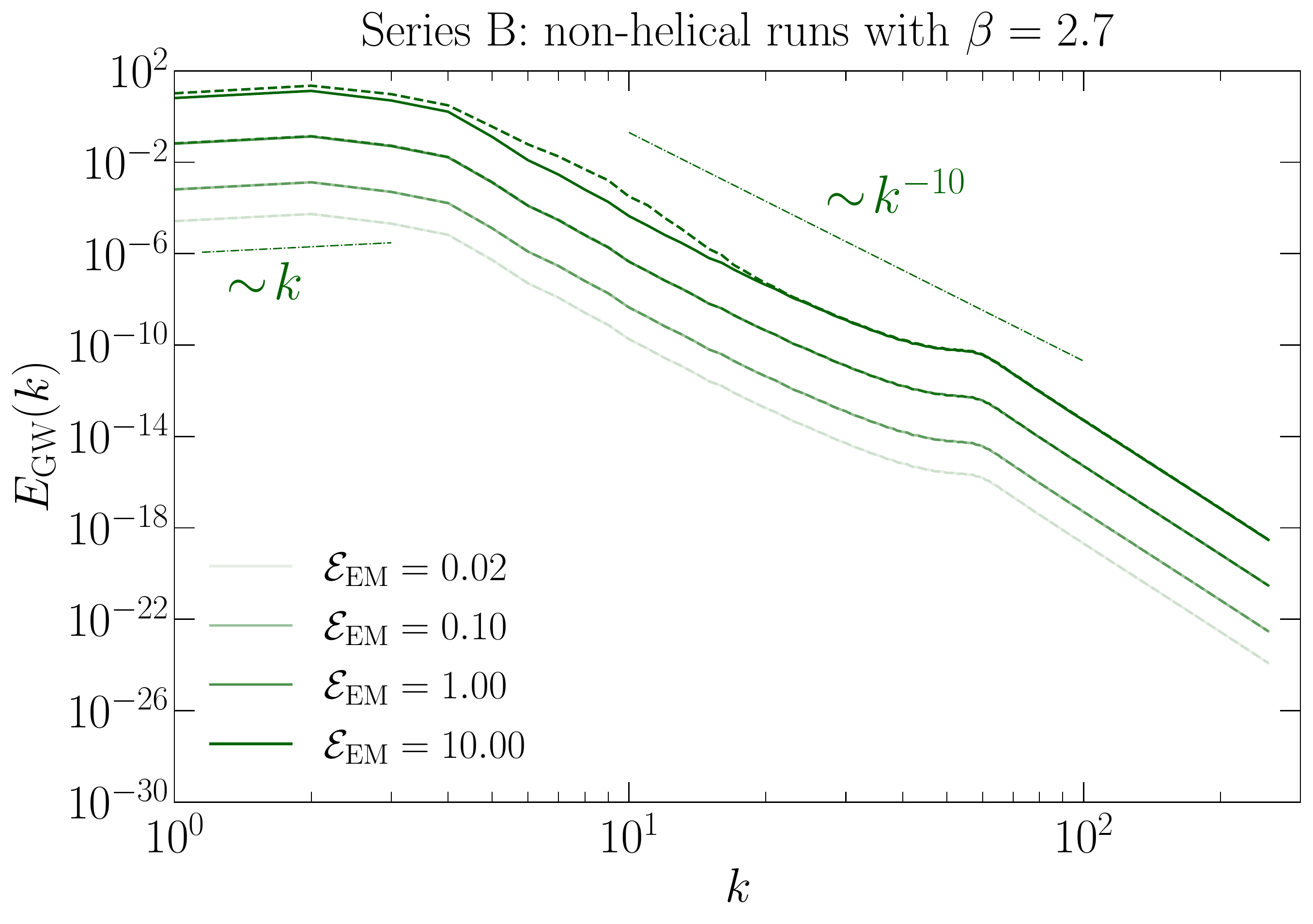}
\includegraphics[width=.40\textwidth]{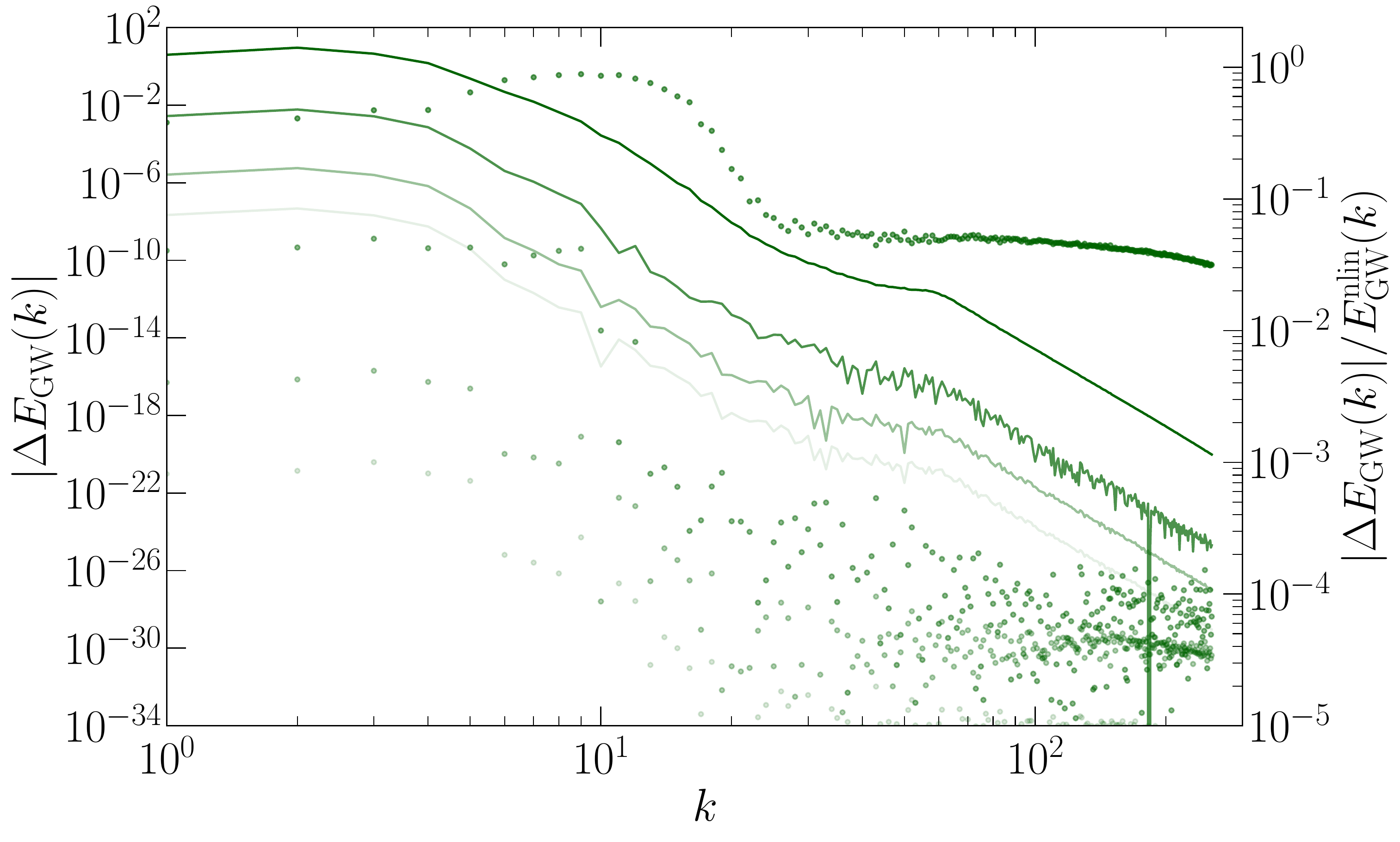}
\includegraphics[width=.36\textwidth]{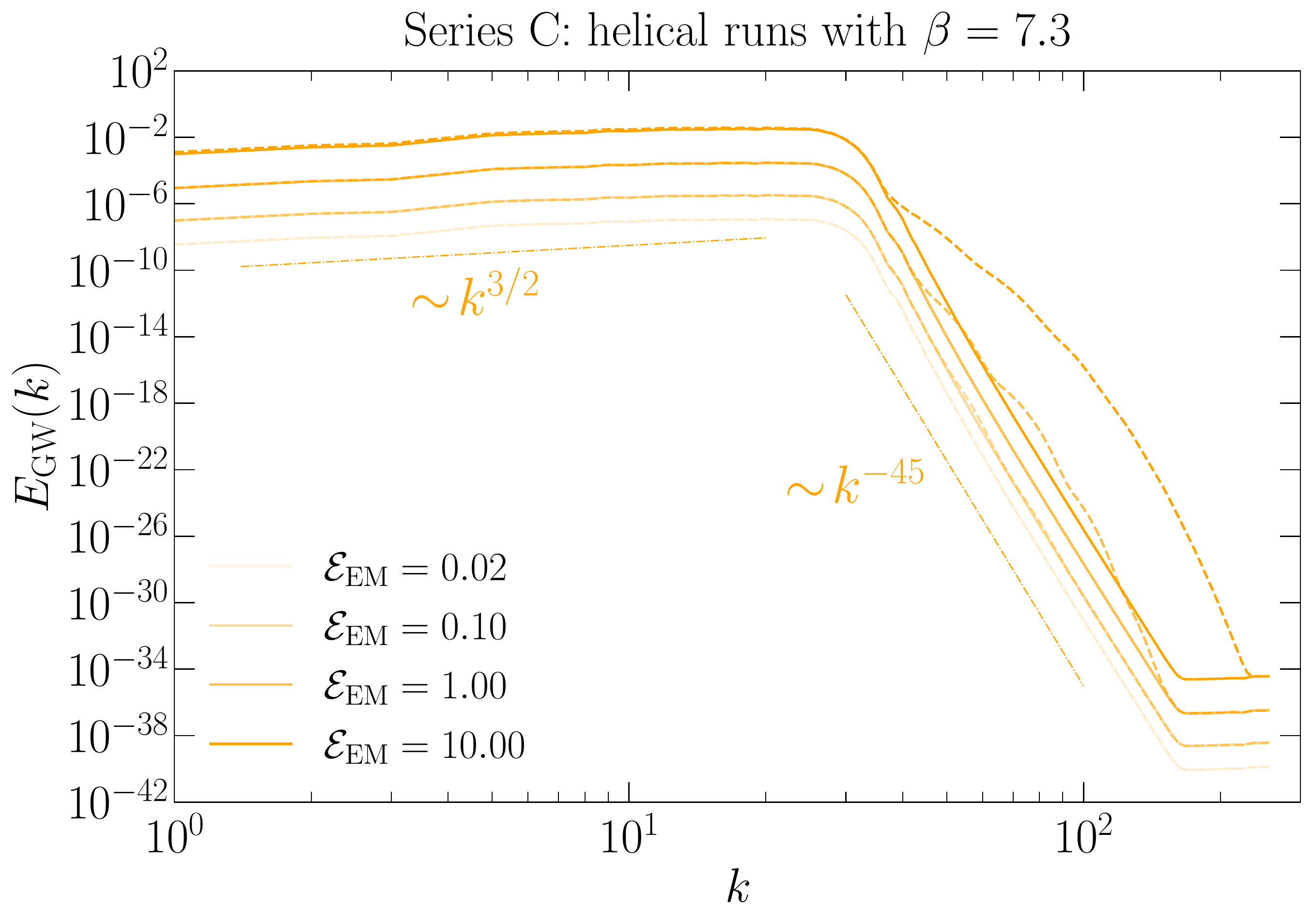}
\includegraphics[width=.40\textwidth]{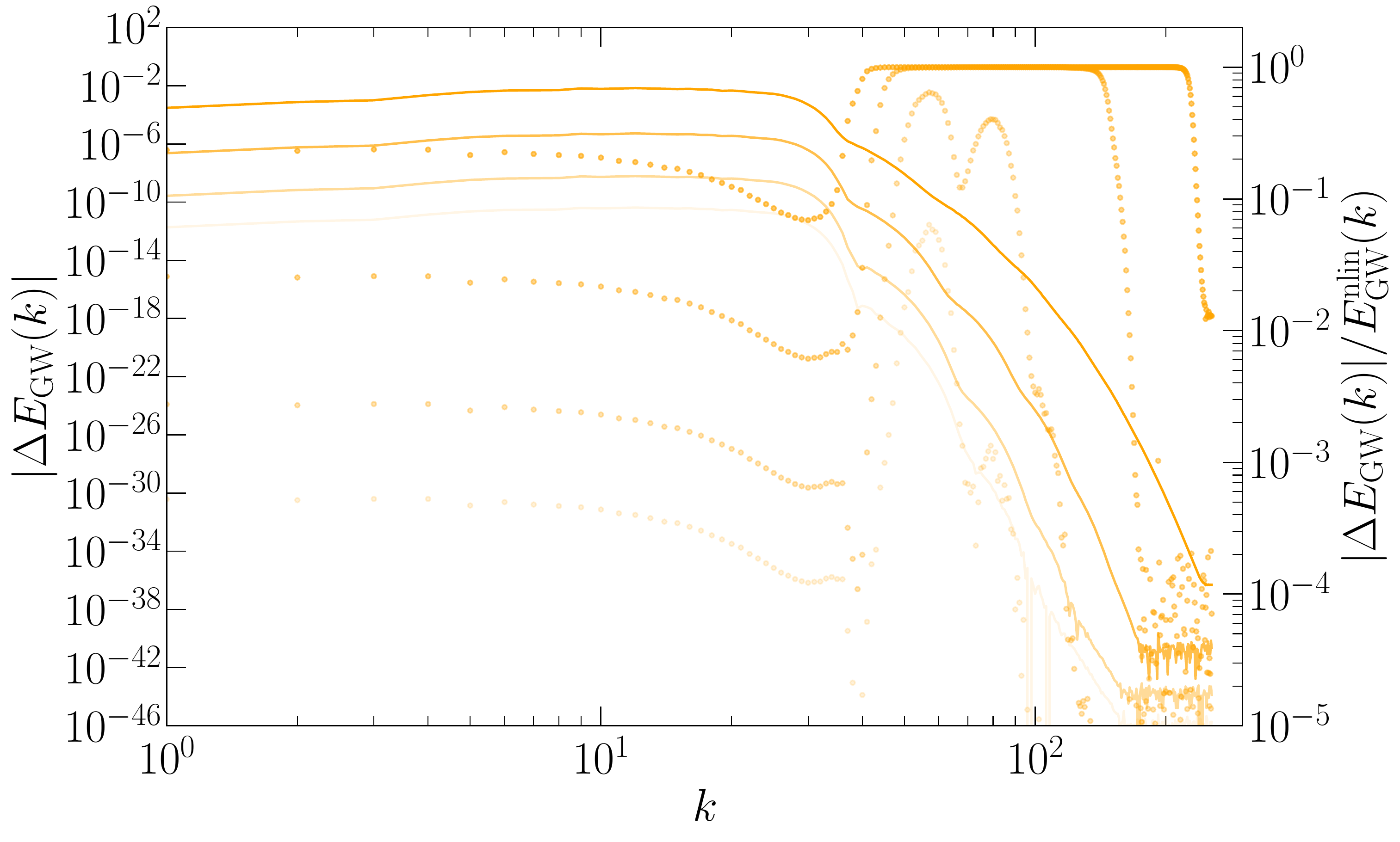}
\includegraphics[width=.36\textwidth]{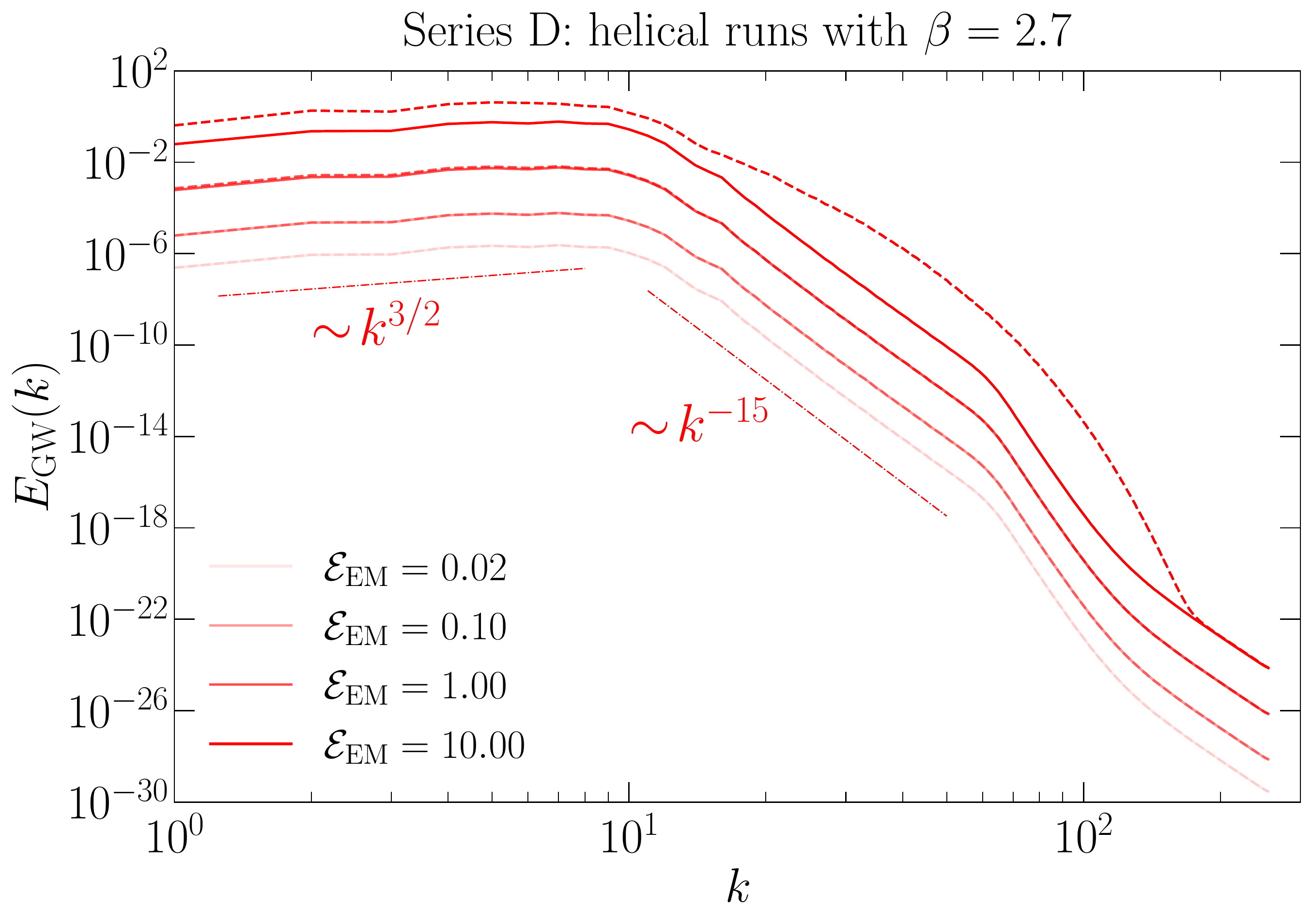}
\includegraphics[width=.40\textwidth]{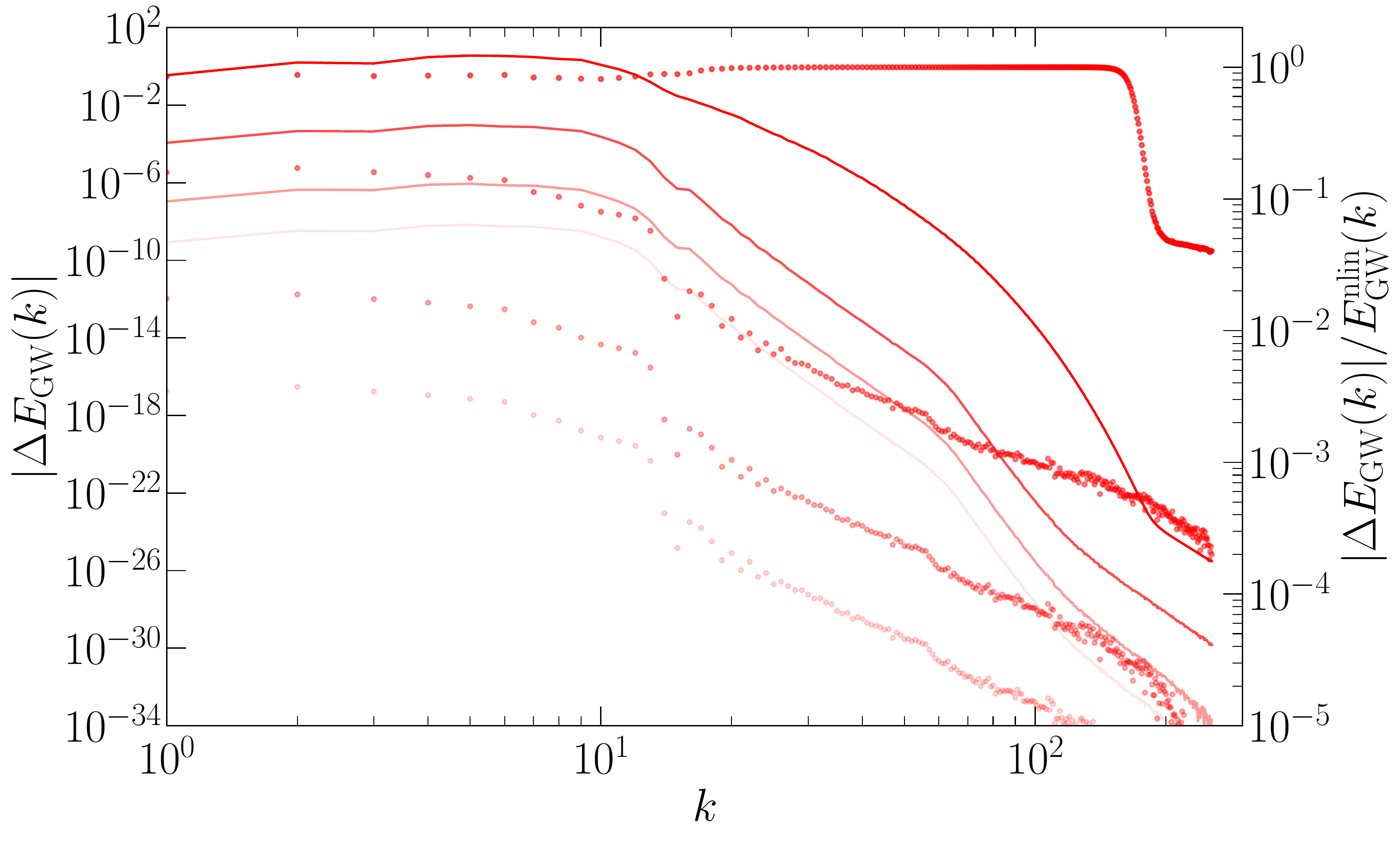}
\includegraphics[width=.36\textwidth]{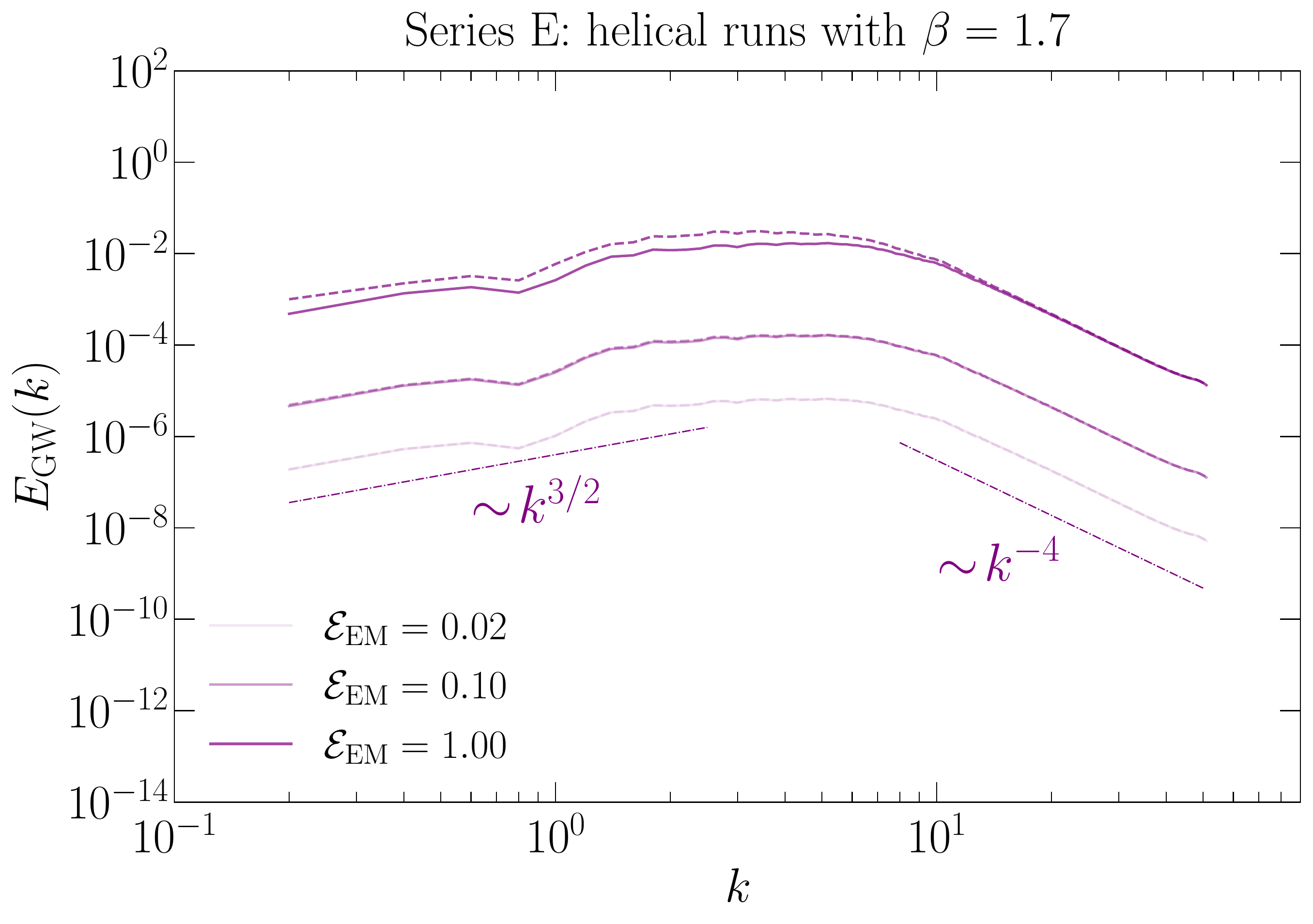}
\includegraphics[width=.40\textwidth]{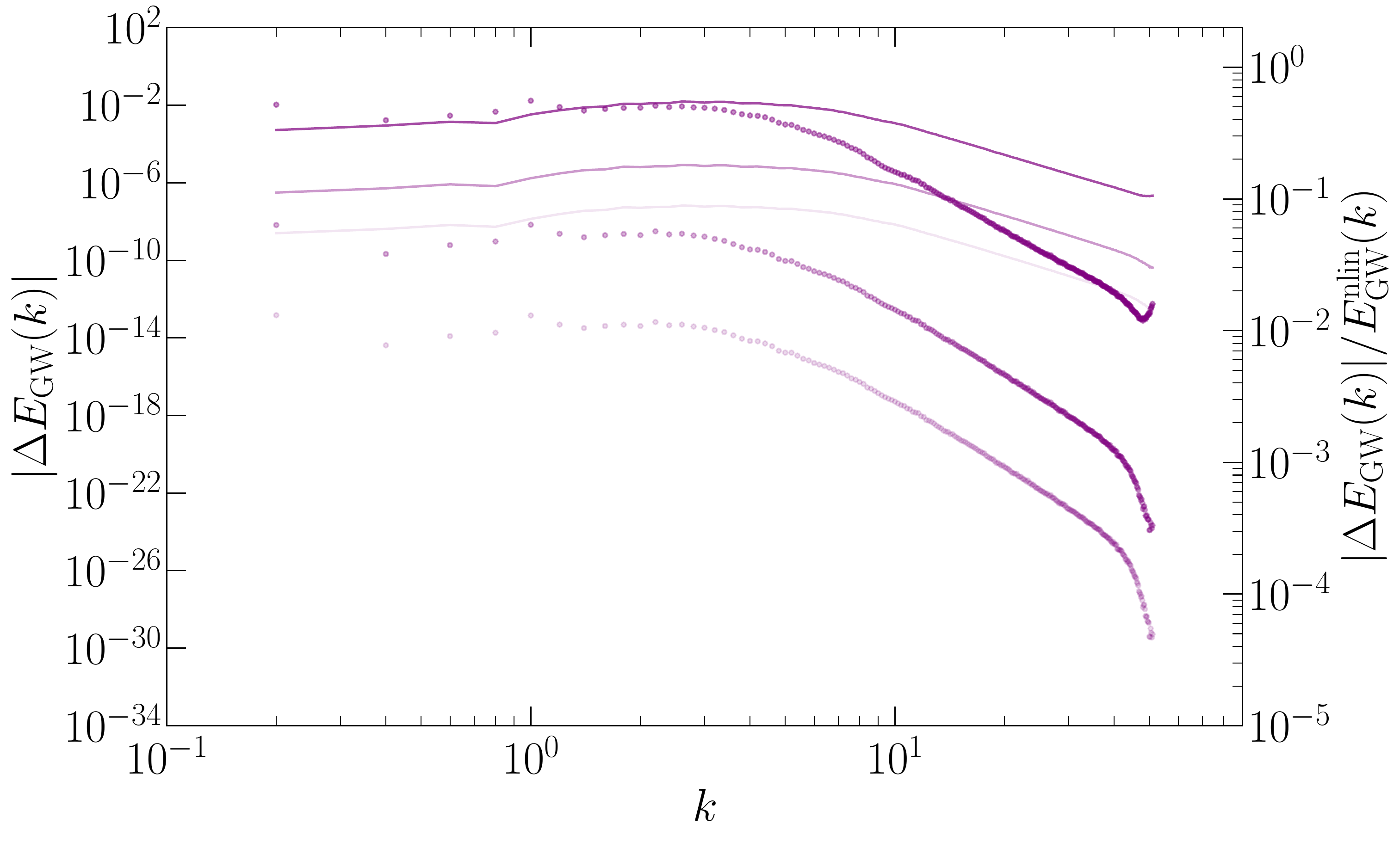}
\caption{Left panels: energy spectra $\EGW(k)$ for $\EEEM = 0.02$
(faintest), 0.1, 1, and 10 (darkest) at $\eta=1$. Linear and nonlinear solutions are in solid and dashed curves, respectively. 
Right panels: differences in the energy spectra $|\Delta\EGW(k)|$ with $y$-axis
to the left and ratio of the two spectra $|\Delta\EGW (k)|/\EGW^\nlin (k)$ in faint 
dotted curves with $y$-axis to the right.}
\label{fig:spec_diffA}
\end{figure*}

To observe the roles of the initial field $B_0$, the reheating parameter $\beta$, and
the helicity $\gamma$, we show the GW energy density spectrum $\EGW (k)$
in Fig.~\ref{fig:spec_diffA}, where each row of panels
corresponds to one of the series A to E.
In the left column, the faintest curves correspond to $\EEEM=0.02$
whereas the darkest shades correspond to the unrealistically extreme case of $\EEEM=10$.
Linear and
nonlinear solutions are shown as solid and dashed curves, respectively.
Figure~\ref{fig:spec_diffA} shows the GW spectra
obtained at $\eta = 1$.

From the left column of Fig.~\ref{fig:spec_diffA}, we see that all linear solutions 
exhibit a shallow spectrum at small $k$, with empirical slopes $\sim \!k$
and $\sim \!k^{3/2}$ for nonhelical and helical cases, respectively.
Around the GW spectral peak, located at $\sim \! 2k_*$, the spectra drop by many orders
of magnitude, although the rate of such sharp drops varies -- it is
quicker for larger values of $\beta$ and for helical cases,
which is in agreement with Ref.~\cite{Brandenburg:2021bfx}.

We then see that all the nonlinear solutions, 
shown as dashed curves in the left column of Fig.~\ref{fig:spec_diffA}, 
closely follow the linear ones, but depart at the steep slopes when the linear spectra start to drop significantly. 
These departures are in the form of an upward ``bulge", indicating that stronger GWs are produced at the nonlinear level. 
The differences diminish towards larger wave numbers. 
Helicity plays a visible role of extending the range of the bulges.
This can be seen by comparing Series A and C in \Fig{fig:spec_diffA},
where the bulges span over only a fraction of
the intermediate $k$ range around $k \sim  10$ without helicity but extend down 
to  the largest wave numbers with helicity.
This is more pronounced in Series C, D, and E, where the top bulge now
hovers above its linear counterpart for almost the entire $k$ range.

In all cases, larger initial $B_0$ produce larger nonlinear effects. 
Since the nonlinear features are not clearly visible in $\EGW(k)$
for all values of $B_0$, we show the difference in GW energy density 
between the linear and nonlinear results, $\Delta \EGW \equiv 
\EGW^\nlin-\EGW$, and the relative difference, $\Delta \EGW/\EGW^\nlin$,
in the right panels of \Fig{fig:spec_diffA}.
We see that the bulges are indeed the features that stand out, as they 
give the largest ratios.
However, whether they contribute significantly to the overall energy difference varies among the series.
In series D and E, as the bulges occur when the linear spectra are shallow,
the nonlinear effects contribute significantly to the overall GW energy density.
On the other hand, spectra for series A, B, and C at small $k$
display no visible bulges,
yet register relatively large values of the difference $\Delta\EGW$.
In these runs, the bulge only appears where the spectra are already
dropping significantly, such that their effect to the total GW energy density is negligible.
At most wave numbers, we observe the nonlinear spectrum to be larger than
the linear one, as expected.
However, at large wave numbers, we find some values of $k$ at which
the opposite occurs,
in particular in the nonhelical series A and B.
We find that the time evolution of the GW modes shows a growing phase followed
by an oscillatory one, as was previously found in other numerical simulations
\cite{RoperPol:2019wvy, Pol:2021uol}.
It is during the oscillatory phase that the linear solution can become
larger than the nonlinear one due to the fact that, even though the latter
has a larger oscillatory amplitude, there is a slight
shift in the phase of the oscillations, allowing the former to acquire larger
values between the minimum and the maximum of the oscillations.

\begin{figure}[t!]
\includegraphics[width=.9\columnwidth]{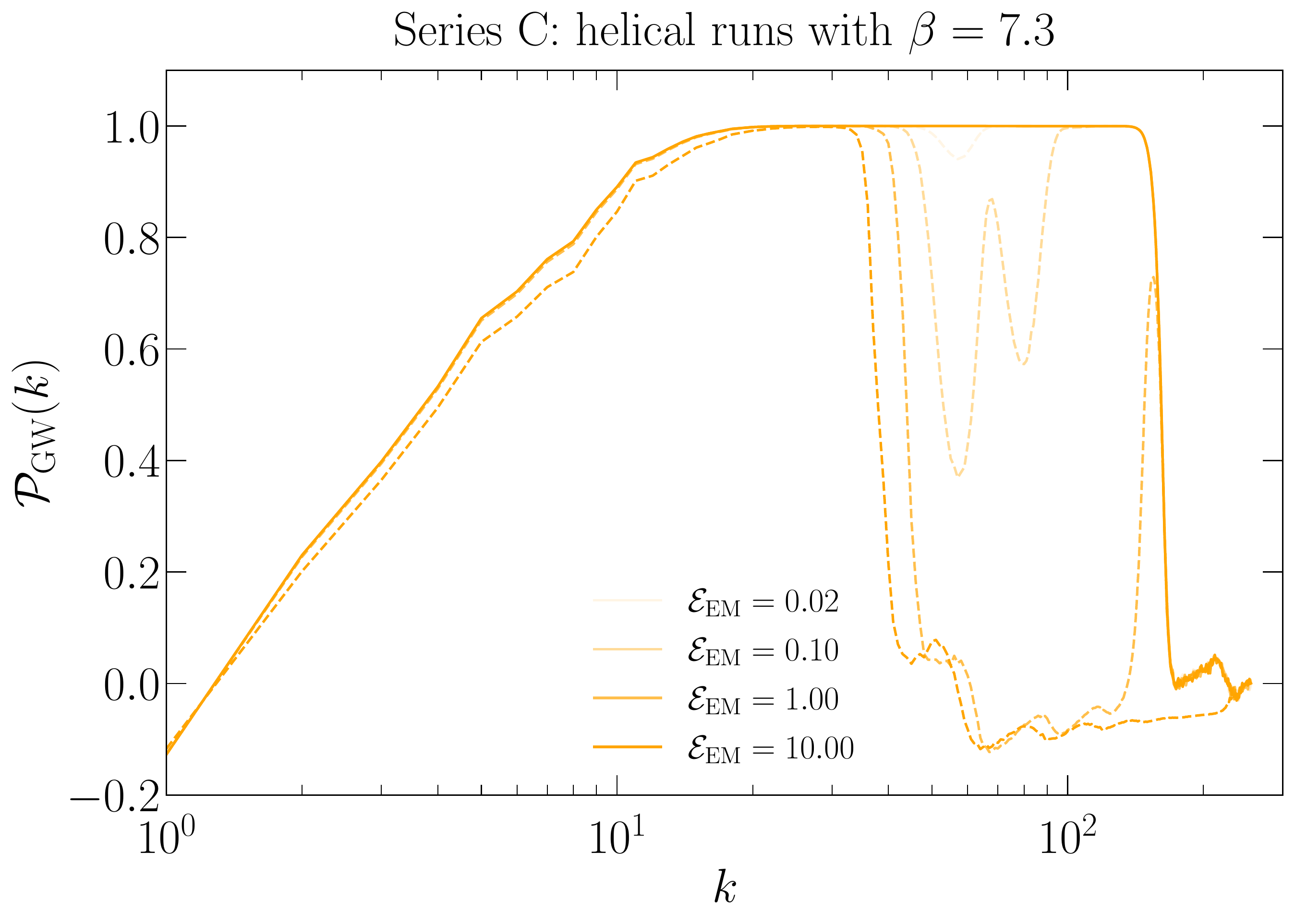}
\includegraphics[width=.9\columnwidth]{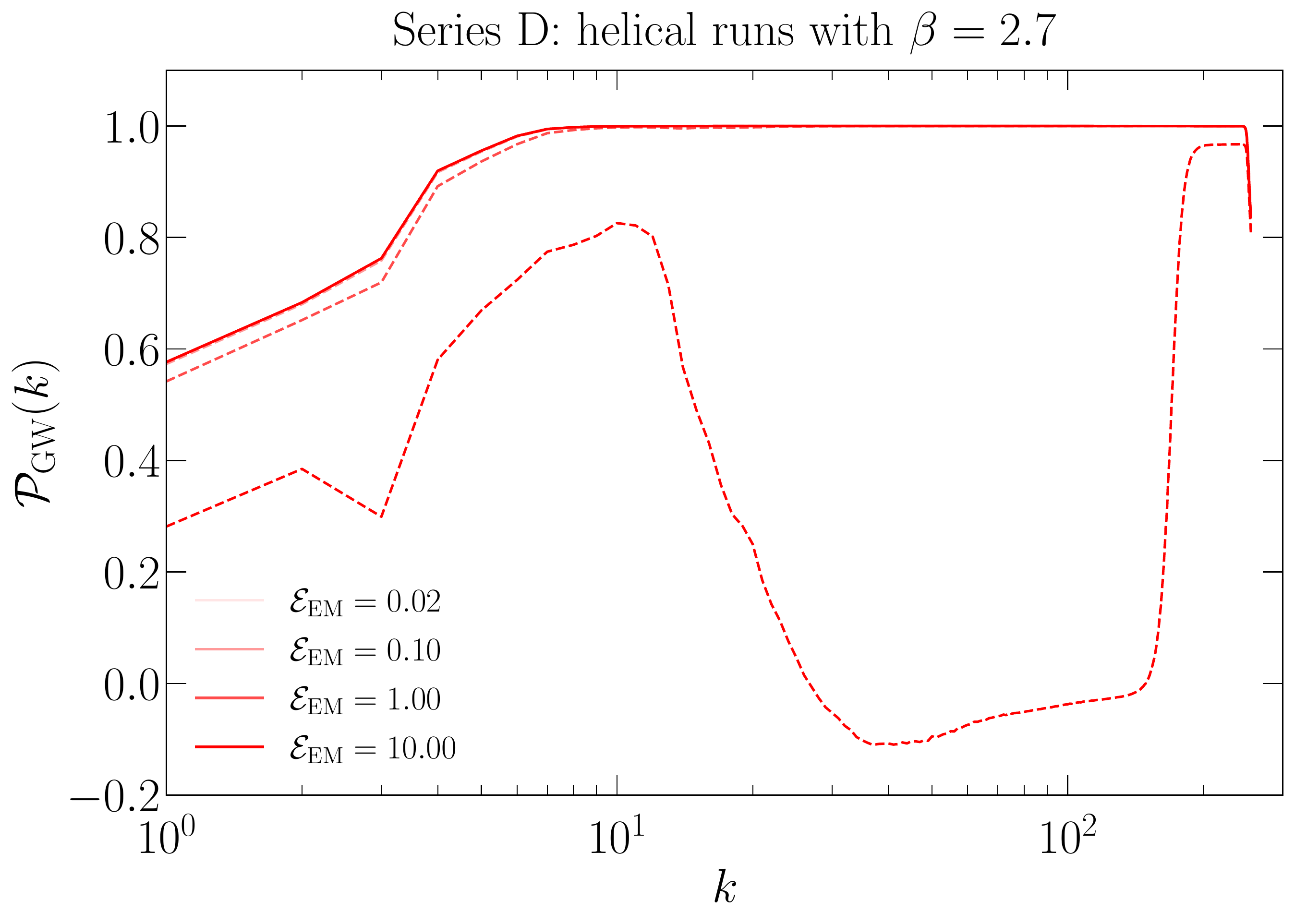}
\includegraphics[width=.9\columnwidth]{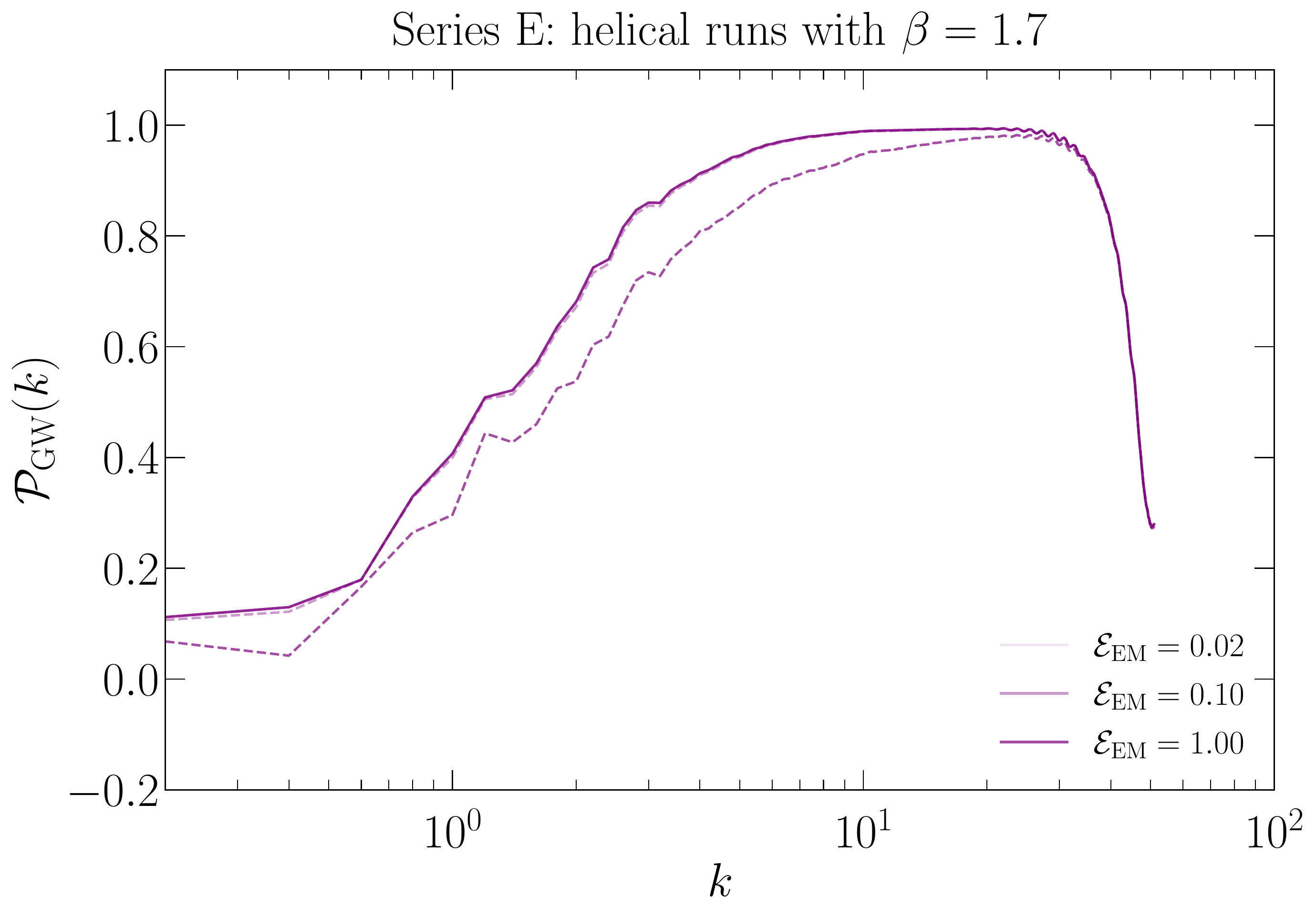}
\caption{GW polarization spectra for $\beta=7.3$ (top panel), 
$\beta=2.7$ (middle panel), and $\beta = 1.7$ (bottom panel)
at $\eta=1$.
Solid and dashed curves are
linear and nonlinear solutions. 
Fainter curves indicate smaller values of $\EEEM$.
In series D and E, the dashed curves for $\EEEM=0.02$, $0.1$, and 1
overlap almost entirely with the solid piece and with each other.}
\label{fig:spec_pol}
\end{figure}

In Fig.~\ref{fig:spec_pol}, we show the degree of circular polarization $\PPGW (k)$,
defined in \Eq{eqn:GWpol}, for
series of helical runs~C, D, and E, with $\beta = 7.3$, $2.7$, and $1.7$, respectively.
We observe that the nonlinearity produces weaker polarization
where the nonlinear strains are larger, which occurs at
large wave numbers, $k \sim 100$,
i.e., where the bulges occur; see
series~C and D in Fig.~\ref{fig:spec_diffA}.
Note that the differences in polarization seem to affect all
variants of series C with $\beta=7.3$, but are almost negligible for
$\beta=2.7$ apart from D4 with $\EEEM=10$.
This is also comparable to the fractional effects $\Delta\EGW(k)/\EGW^\nlin(k)$ shown in series C and D of Fig.~\ref{fig:spec_diffA},
where all variants of series C register significant nonlinear effects,
but among series D only D4 does.
Similarly to series D, the runs of series E do not show a large decrease
in the polarization, only visible for E3, which corresponds to the largest
EM energy density considered in these runs, i.e., $\EEEM = 1$.
This is expected since the decrease in polarization is
induced by the increased ratio in the GW spectrum as shown in \Eq{PGW_order}.

\subsection{Scaling with $\EEEM$}
\label{ssec:dependence_on_EEEM}

In Eq.~\eqref{eqn:OME_dEEGW}, we have alluded to the expected
order-of-magnitude relation between $\Delta\EEGW$ and $\EEEM$.
Here we present the empirical findings of such a relation.

\begin{figure}[t]
\includegraphics[width=.9\columnwidth]{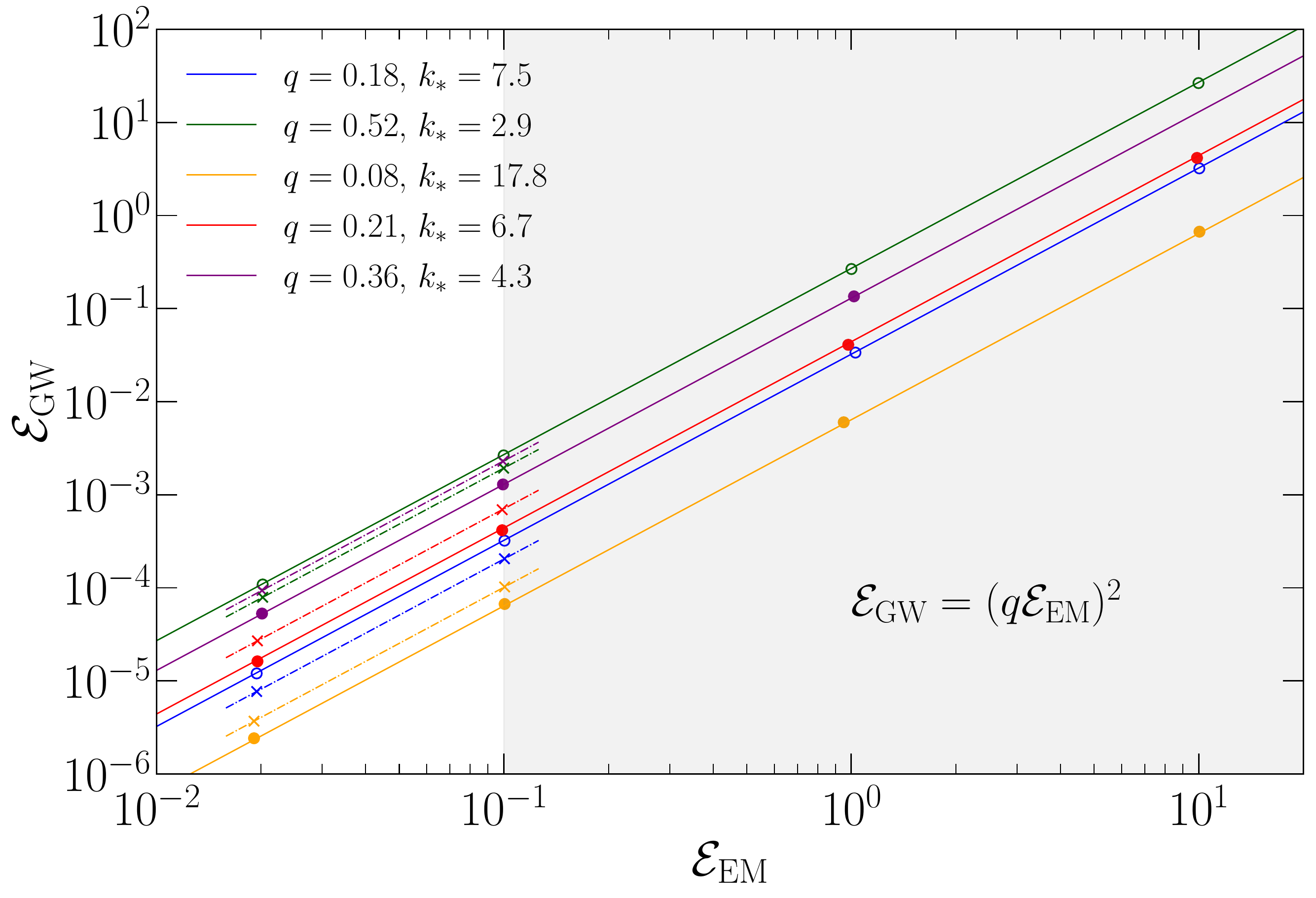}
\includegraphics[width=.9\columnwidth]{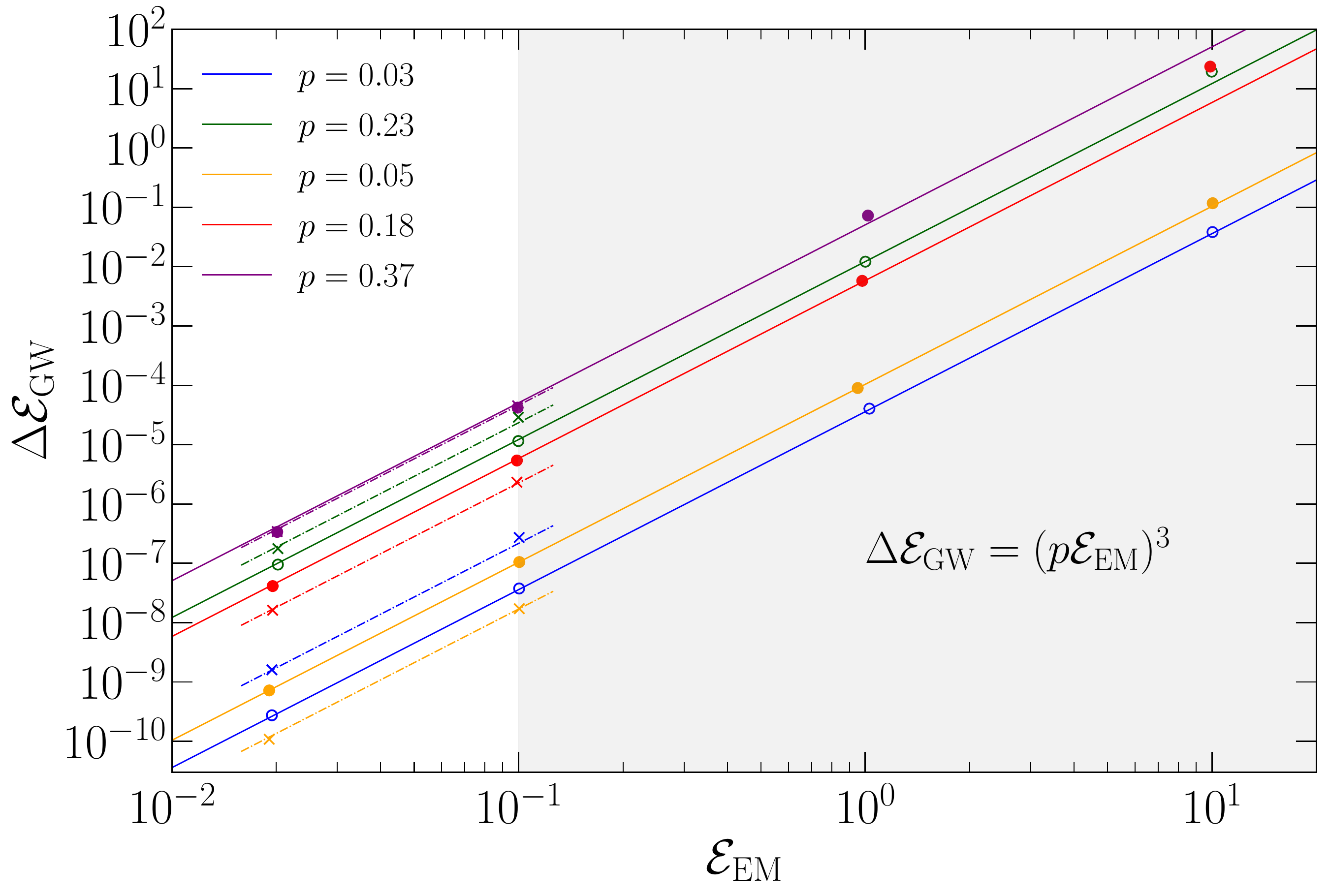}
\includegraphics[width=.9\columnwidth]{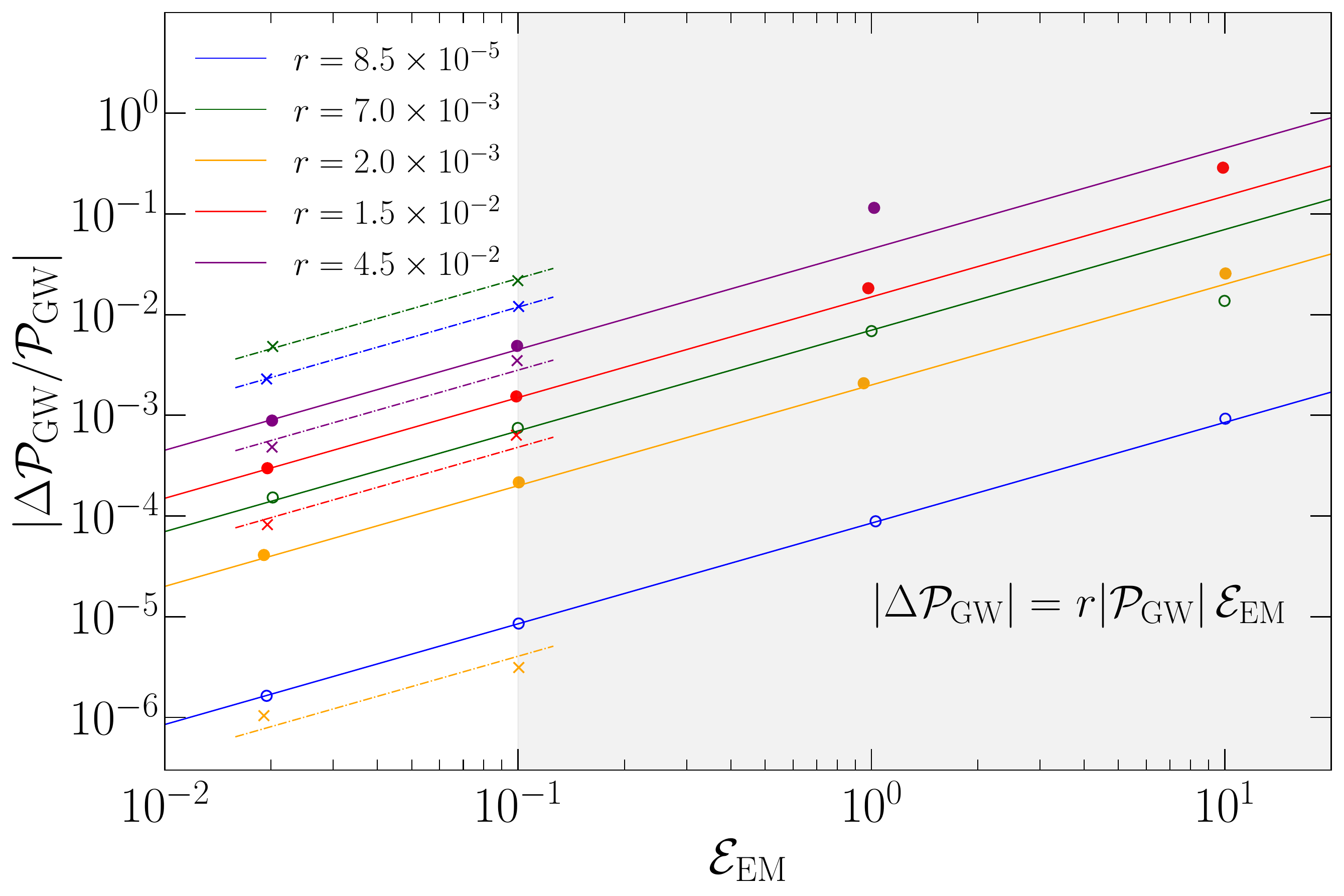}
\caption{GW energy density $\EEGW$ (top panel), energy difference
$\Delta\EEGW$ (middle panel), and relative polarization suppression
$|\Delta \PPGW/\PPGW|$ (bottom panel) against sourcing energy density $\EEEM$
at the end of reheating.
We show series A (blue hollow circles), series B (green hollow circles),
series C (yellow dots), series D (red dots), and series E (purple dots);
see \Tab{tab:runs}.
Dashed lines and crosses are the fittings corresponding
to the saturated values of the extended runs A', B', C', D', and E'.
Shaded areas indicate $\EEEM>0.1$.}
\label{fig:comp1}
\end{figure}

We show in \Fig{fig:comp1} the GW production efficiency by comparing the
GW energy reached at the end of reheating $\eta_*$,
$\EEGW$ (top panel), and the energy difference,
$\Delta\EEGW$ (middle panel), against the maximum
sourcing energy, also reached at $\eta_*$, $\EEEM$.
The gray shaded regions in both panels correspond to values of EM energy 
densities larger than the upper bound imposed by BBN
\cite{Shvartsman:1969mm,Grasso:1996kk,Kahniashvili:2009qi}.
We find that the quadratic relation $\EEGW \propto \EEEM^2$ and the cubic relation $\Delta\EEGW \propto \EEEM^3$ hold at all values of $\EEEM$ considered.

Then, we denote the efficiency coefficients as $q$ and $p$ and rewrite
the order-of-magnitude expressions in Eqs.~\eqref{eqn:OME_EEGW} and
\eqref{eqn:OME_dEEGW} as
\begin{equation}
\EEGW = (q\EEEM)^2,\;
\Delta\EEGW = (p\EEEM)^3.
\label{eqn:EEGW_q2p3}
\end{equation}
We obtain the pairs of values $(q, p)$ that fit the runs of \Tab{tab:runs}
for each of the series A to E.
These are shown in Table~\ref{tab:q_p_r} and in the legend of \Fig{fig:comp1}.
We see that runs with $\beta=2.7$ overall produce GWs more efficiently than those with $\beta=7.3$, with or without helicity.
The helical runs with $\beta = 1.7$ have even larger efficiency than
the other helical runs.
Hence, smaller values of $\beta$ seem to induce GWs with larger efficiencies.

\begin{table}[b!]
\caption{Empirical findings of the coefficients $(q,p)$, $(\tilde q,\tilde p)$, $r$, and $\tilde r$
for the runs of \Tab{tab:runs}.
A, B, C, D, and E refer to the values at the end of reheating
while A', B', C', D', and E' refer to the saturated values.
}
\centering
\renewcommand{\arraystretch}{1.25}
\begin{tabular}{l|cc|rr}
Runs & $(q,p)$ & $(\tilde q,\tilde p)$ & $r$ \hspace{6.5mm} & $\tilde r$ \hspace{6.5mm} \\\hline
A & $(0.18, 0.03)$ & $(1.36, 0.25)$ &
$8.5 \times 10^{-5}$ & $6.4 \times 10^{-4}$ \\
\blue{A'} & \blue{$(0.14, 0.06)$} & \blue{$(1.08, 0.45)$} &
\blue{$1.2 \times 10^{-1}$} & \blue{$9.0 \times 10^{-1}$} \\
B & $(0.52, 0.23)$ & $(1.53, 0.68)$ &
$7.0 \times 10^{-3}$ & $2.1 \times 10^{-2}$ \\
\green{B'} & \green{$(0.44, 0.29)$} & \green{$(1.29, 0.84)$} &
\green{$2.3 \times 10^{-1}$} & \green{$6.7 \times 10^{-1}$} \\\hline
C & $(0.08, 0.05)$ & $(1.42, 0.84)$ &
$2.0 \times 10^{-3}$ & $3.6 \times 10^{-2}$ \\
\orange{C'} & \orange{$(0.10, 0.03)$} & \orange{$(1.79, 0.46)$} &
\orange{$4.0 \times 10^{-5}$} & \orange{$7.2 \times 10^{-4}$} \\
D & $(0.21, 0.18)$ & $(1.41, 1.20)$ &
$1.5 \times 10^{-2}$ & $1.0 \times 10^{-1}$ \\
\red{D'} & \red{$(0.27, 0.13)$} & \red{$(1.78, 0.88)$} &
\red{$4.8 \times 10^{-3}$} & \red{$3.2 \times 10^{-2}$} \\
E & $(0.36, 0.37)$ & $(1.54, 1.58)$ &
$4.5 \times 10^{-2}$ & $1.9 \times 10^{-1}$ \\
\purple{E'} & \purple{$(0.48, 0.36)$} & \purple{$(2.06, 1.53)$} &
\purple{$2.8 \times 10^{-2}$} & \purple{$1.2 \times 10^{-1}$} \\
\end{tabular}
\label{tab:q_p_r}
\end{table}

In some earlier parameterizations of GW efficiencies
\cite{RoperPol:2019wvy,Brandenburg:2021xyz,Pol:2021uol}, it was found advantageous
to take the dependence on the typical wave number $k_*$ of Eq.~\eqref{eqn:kstar}
into account
via $\tilde{q}\equiv q k_*$.
Analogously, we might write $\tilde{p}=k_* p$.
Then Eq.~\eqref{eqn:EEGW_q2p3} is rewritten as
\begin{equation}
\EEGW = (\tilde{q}\EEEM/k_*)^2,\;
\Delta\EEGW = (\tilde{p}\EEEM/k_*)^3.
\label{eqn:EEGW_q2p3new}
\end{equation}
The corresponding pairs $(\tilde{q}, \tilde{p})$
are also listed in Table~\ref{tab:q_p_r}.
We see that, for both nonhelical and helical cases, 
$\tilde{q}$ and $\tilde{p}$ are indeed justified parameterizations,
as they vary less across different values of $\beta$ than $q$ and $p$.

In an analogous manner to the dependences of $\EEGW$ and $\Delta \EEGW$
on $\EEEM$, we also found in \Eq{PGW_order} an expected linear dependence
of the relative ratio of the decrease in polarization due to the nonlinear contributions, i.e., $\Delta \PPGW/\PPGW \propto \EEEM$.
This relation is also found in the numerical simulations; see bottom panel of
\Fig{fig:comp1}, especially for lower values of $\EEEM$.
In this case, we define the polarization suppression coefficients $r$ and $\tilde r \equiv r k_*$ such that
\begin{equation}
   \Big|\frac{\Delta \PPGW}{\PPGW}\Big| = r\, \EEEM, \quad
   \Big|\frac{\Delta \PPGW}{\PPGW}\Big| = \tilde r \,
   \EEEM/k_*,
\end{equation}
and compute their values by fitting the results from the simulations,
listed in \Tab{tab:runs}.
The resulting values are shown in \Tab{tab:q_p_r}.
Here $\tilde r$ is also a somewhat better choice of parameterization than $r$.

\subsection{GW evolution after the end of reheating}
\label{ssec:late_evol}

\begin{figure}[t]
\includegraphics[width=\columnwidth]{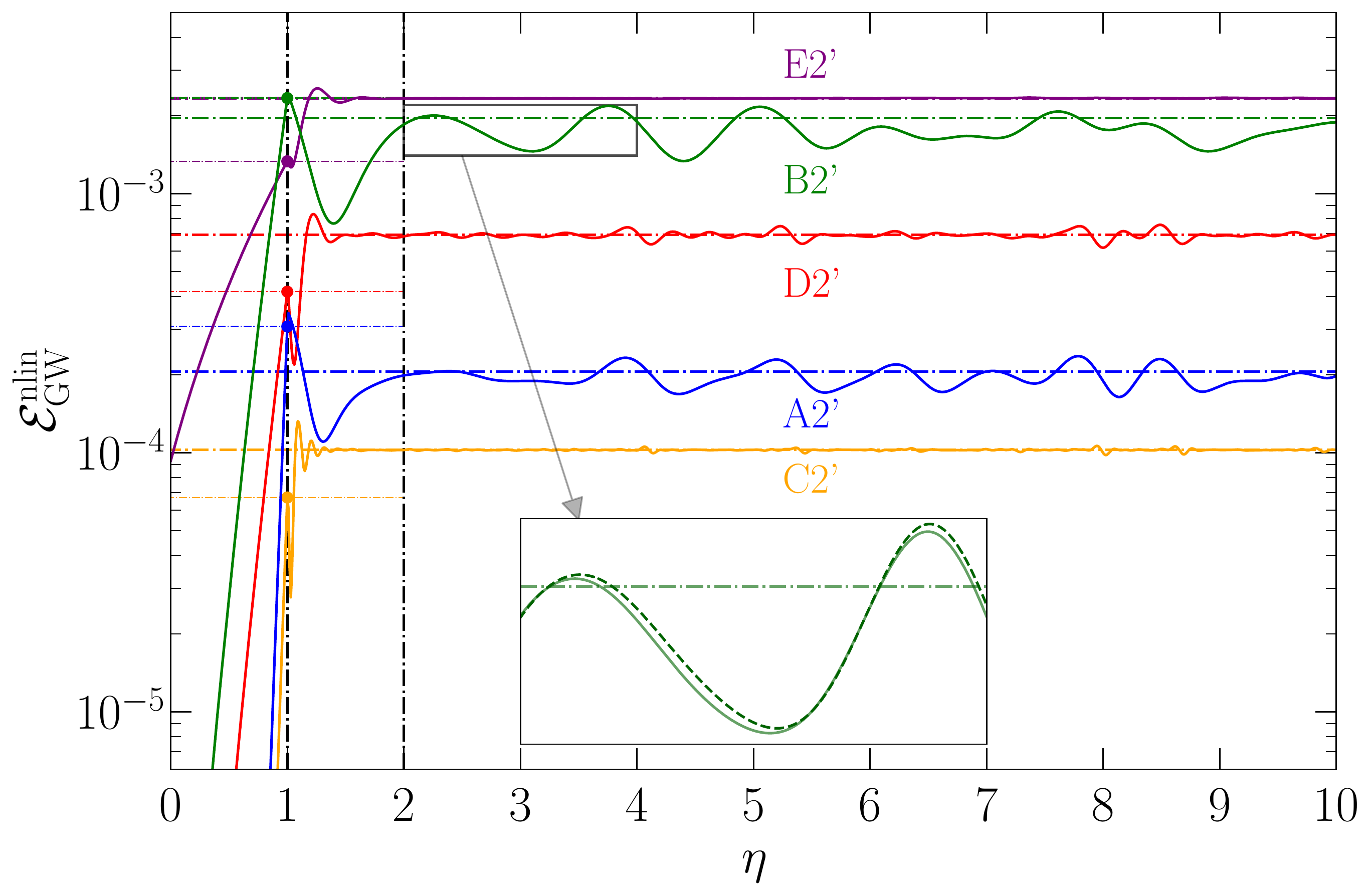}
\includegraphics[width=\columnwidth]{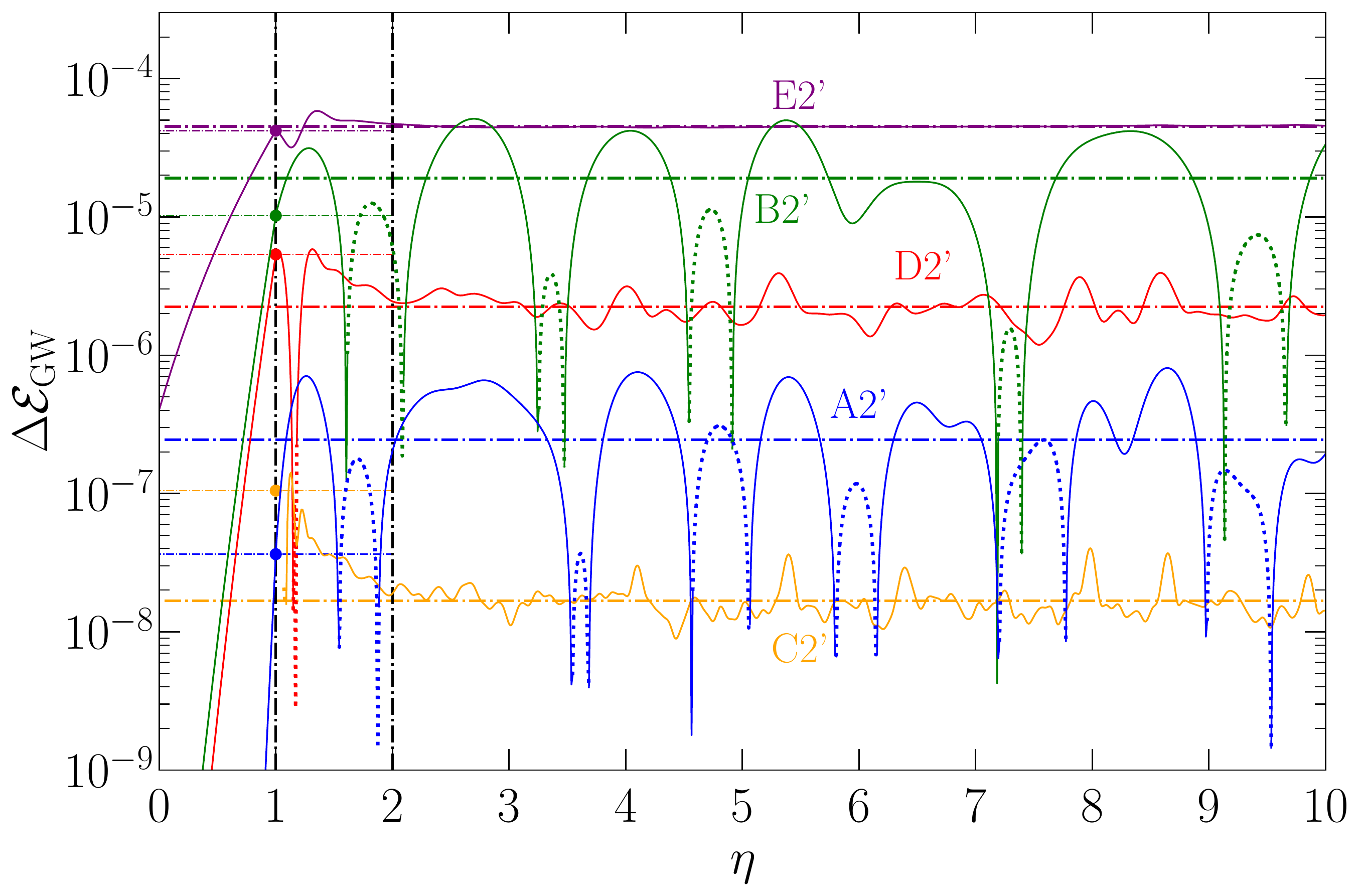}
\caption{Time evolution of the nonlinear GW energy density (upper panel) and
its difference with the linear one (bottom panel). We show runs~A2' (blue), 
B2' (green), C2' (orange), D2' (red), and E2' (purple), all corresponding to
$\EEEM = 0.1$.
The horizontal lines are the averaged values over times larger than $\eta = 2$
and the values at $\eta = 1$.
The vertical lines correspond to $\eta = 1$, i.e.,
the end of reheating, and $\eta = 2$.
The zoomed-in plot (upper panel) shows the difference between nonlinear (dashed) and linear solution for one of the runs.
Positive and negative values in the difference (lower panel) are shown using solid and dotted lines, respectively, in the nonhelical runs, while helical runs only present positive values.}
\label{fig:ts_overshoot}
\end{figure}
The long-term time evolution of GWs, in the absence of the EM source
are  studied with runs~A', B', C', D', and E'.
The results for the runs with $\EEEM = 0.1$
can be seen in Fig.~\ref{fig:ts_overshoot},
where both $\EEGW^\nlin$ (top) and $\Delta\EEGW$ (bottom) become oscillatory around a stationary value\footnote{We could analogously refer to
the linear energy density $\EEGW$ since the difference between the nonlinear and linear is at least 1.5 orders of magnitude smaller.} after $\eta\sim2$.
The resulting values of averaging over oscillations (i.e., between $\eta=2$
and 10) are recorded as the saturated values in Table~\ref{tab:runs}.
Additionally, the parameterization coefficients are obtained analogously and shown in
Table~\ref{tab:q_p_r}.
We observe that the saturated values of the nonlinear GW energy density
become slightly larger than those at the end of reheating in the helical
cases, while the opposite is found in the nonhelical runs.
On the other hand, the difference between nonlinear and linear solutions
becomes larger for nonhelical runs and smaller for helical ones.
This is probably related to the fact that the nonhelical runs present larger
oscillations of the solutions over time already in the stationary phase.
As discussed in \Sec{ssec:dependence_on_beta}, the linear solution might become
larger than the nonlinear one at some exceptional times due to a different phase
in the time oscillations of the GW modes, even though the latter presents larger
oscillatory amplitudes.
This behavior can similarly occur in the total GW energy density,
integrated over wave numbers, as it is seen in \Fig{fig:ts_overshoot} (see
zoomed-in plot of the upper panel).
We observe that between the minimum and maximum of the oscillations, due
to the shift in phase, the nonlinear solution is retarded in time, allowing
smaller values with respect to the linear one.
The negative values of the differences (shown in the lower panel) always occur
at this moment of the oscillations, but the opposite is not
necessarily true.

\begin{figure}[t]
\includegraphics[width=\columnwidth]{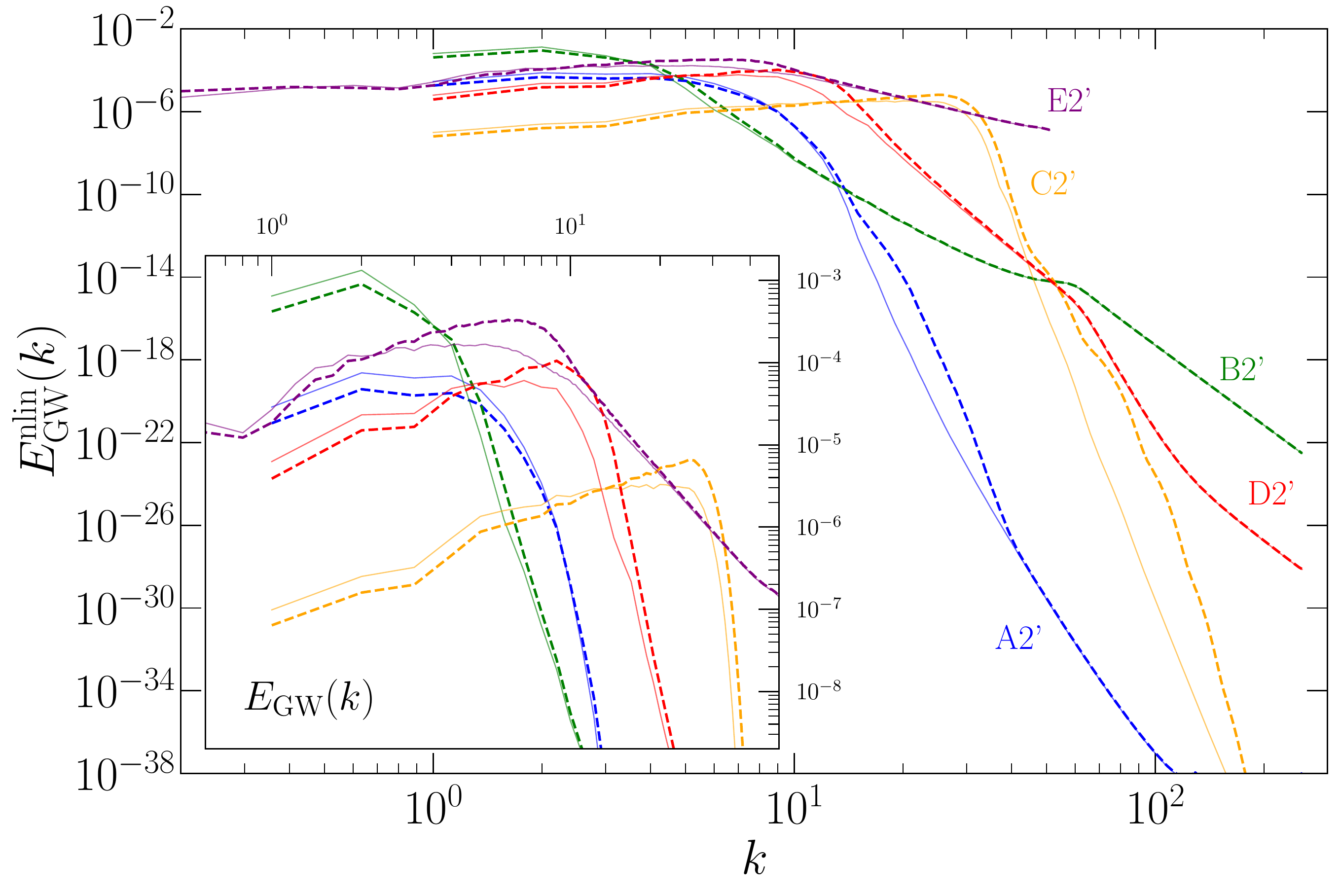}
\caption{Nonlinear GW energy spectra. We show runs~A2' (blue), 
B2' (green), C2' (orange), D2' (red), and E2' (purple), all corresponding to
$\EEEM = 0.1$.
The solid thin lines correspond to the spectra at the end of reheating,
while the dashed lines to the saturated values of the spectra.
The inset shows the linear GW spectra near the spectral peak, where
both linear and nonlinear solutions are almost identical.}
\label{fig:EGW_overshoot}
\end{figure}

Besides the total GW energy density, the individual GW modes also enter an
oscillatory phase around stationary values.
In \Fig{fig:EGW_overshoot}, we show the linear and nonlinear spectra, comparing those obtained
at $\eta = 1$ with the saturated values.
In all cases, the GW spectra decrease after $\eta = 1$ at low wave numbers,
while they increase around the spectral peak.
The latter increase is less pronounced in the nonhelical runs, which then explains
why in these runs, the total saturated GW energy density becomes smaller than
its value at $\eta = 1$.
In the helical runs, the increase around the spectral peak compensates for 
the decrease at low wave numbers, hence leading to a total increase of the
saturated values.

The differences between the linear and nonlinear GW spectra are not noticeable
by eye at $\eta = 1$; see runs with $\EEEM = 0.1$ in \Fig{fig:spec_diffA}.
However, the saturated values of the spectra show a pronounced bulge at large
wave numbers for the nonlinear solutions of runs A2' and C2', i.e., the runs
with $\beta = 7.3$.
Since this feature was observed in \Fig{fig:spec_diffA} for larger values of
$\EEEM$, we can imply that it is enhanced by letting the GW equation evolve
for longer times after the end of reheating.

\begin{figure}[t]
\includegraphics[width=\columnwidth]{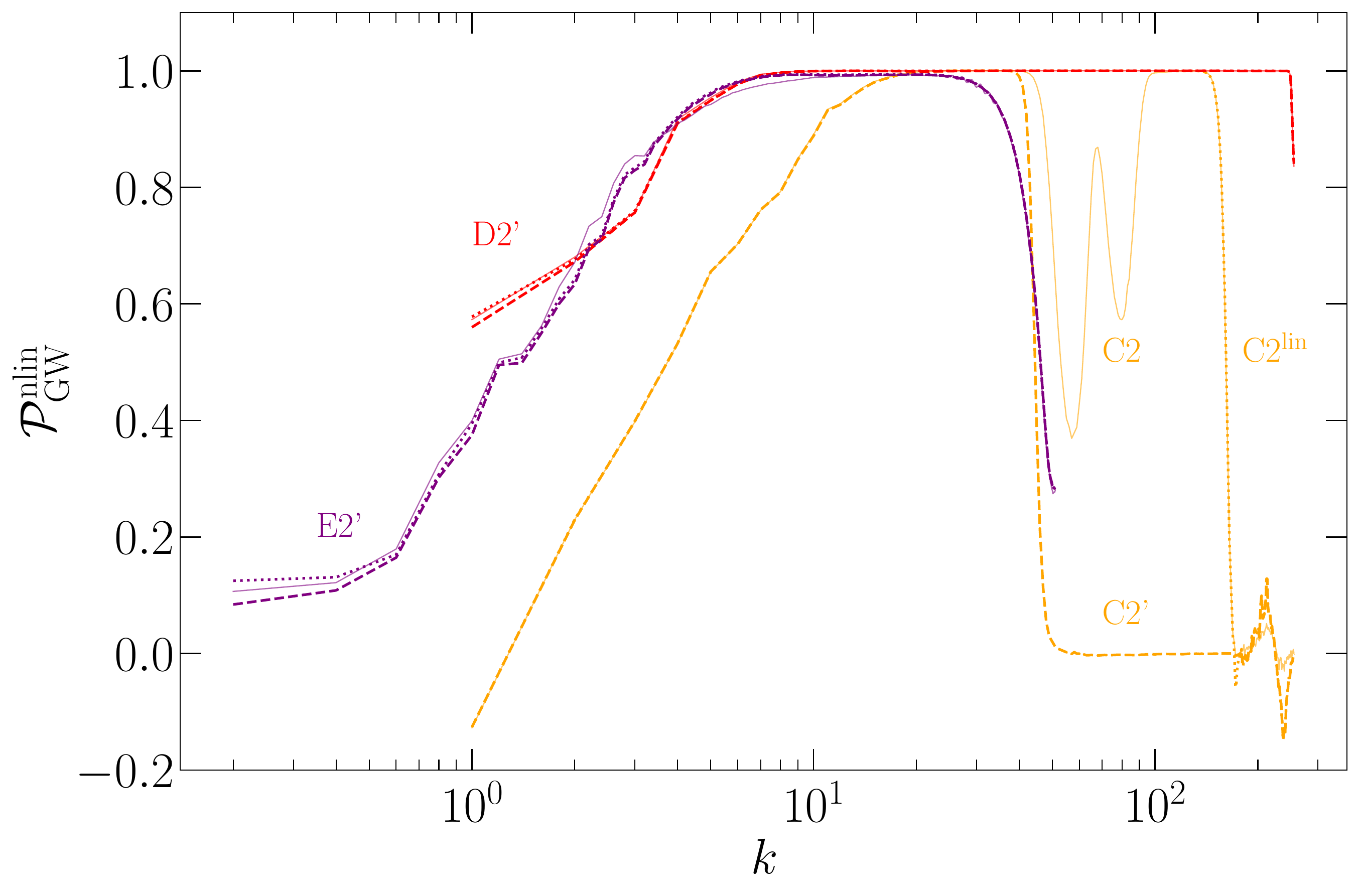}
\caption{GW polarization spectra. We show helical runs
C2' (orange), D2' (red), and E2' (purple), all corresponding to
$\EEEM = 0.1$.
We compare the linear (dotted) and nonlinear (dashed) saturated polarization
spectra with those at the end of reheating (thin solid curves).}
\label{fig:PGW_overshoot}
\end{figure}

The polarization spectra of the helical runs is shown in \Fig{fig:PGW_overshoot}.
As in \Fig{fig:spec_pol}, series D and E do not show a significant difference between linear
and nonlinear polarizations for $\EEEM = 0.1$.
A noticeable difference appears in series C, which already showed a strong
decrease in polarization at the end of reheating, from 1 down to values below 0.4
at wave numbers between 50 and 150, which corresponds to the position of the
GW energy density bulge.
In this case, the saturated polarization shows an even stronger decrease, going all
the way down to zero, i.e., becoming unpolarized, in the same range of wave numbers.
This again coincides with the appearance of a stronger bulge in the GW energy
density; see \Fig{fig:EGW_overshoot}.

\subsection{Comparison with observational limits}
\label{ssec:observation}
\begin{figure*}[t!]
\includegraphics[width=\columnwidth]{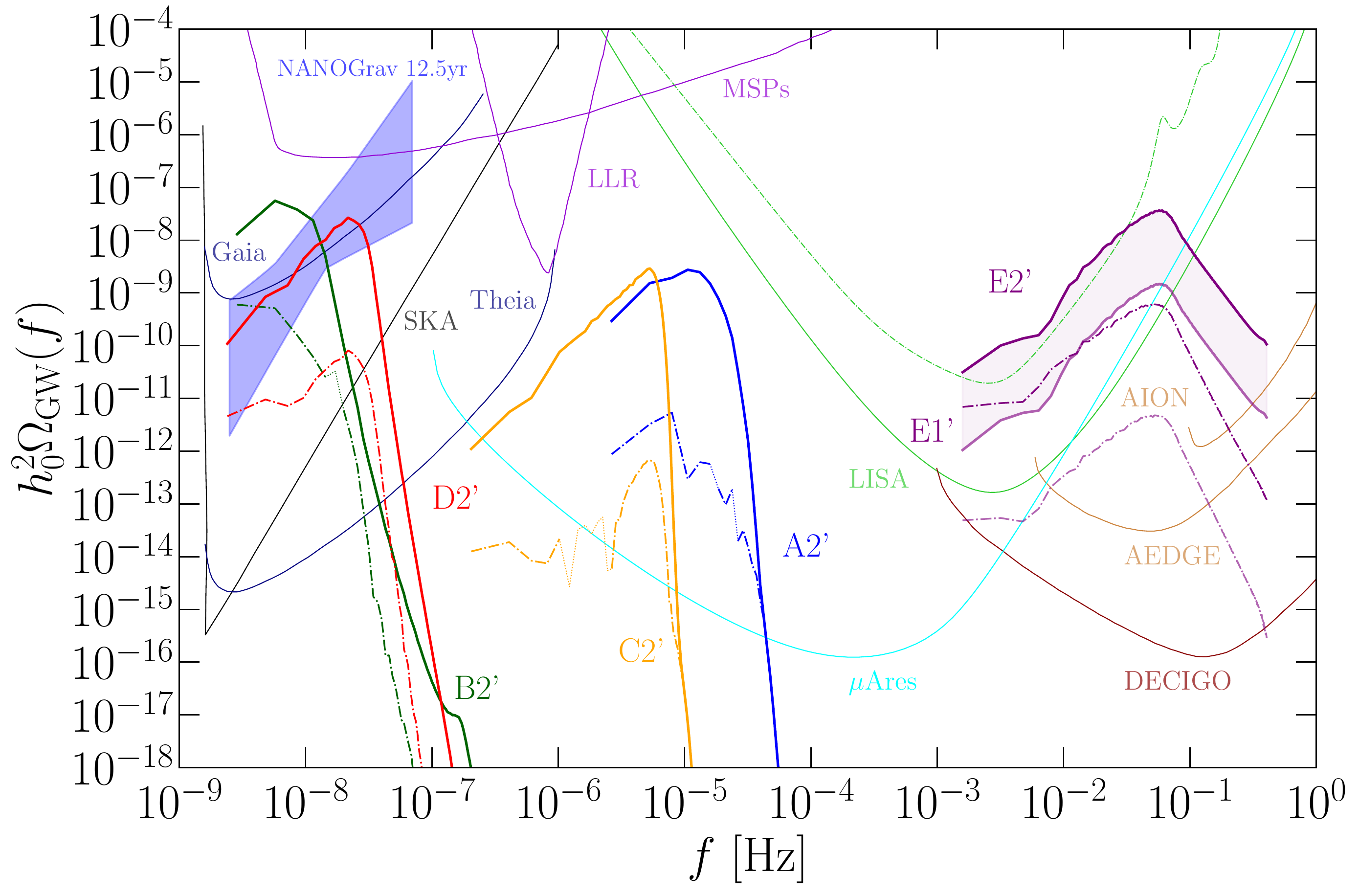}
\includegraphics[width=\columnwidth]{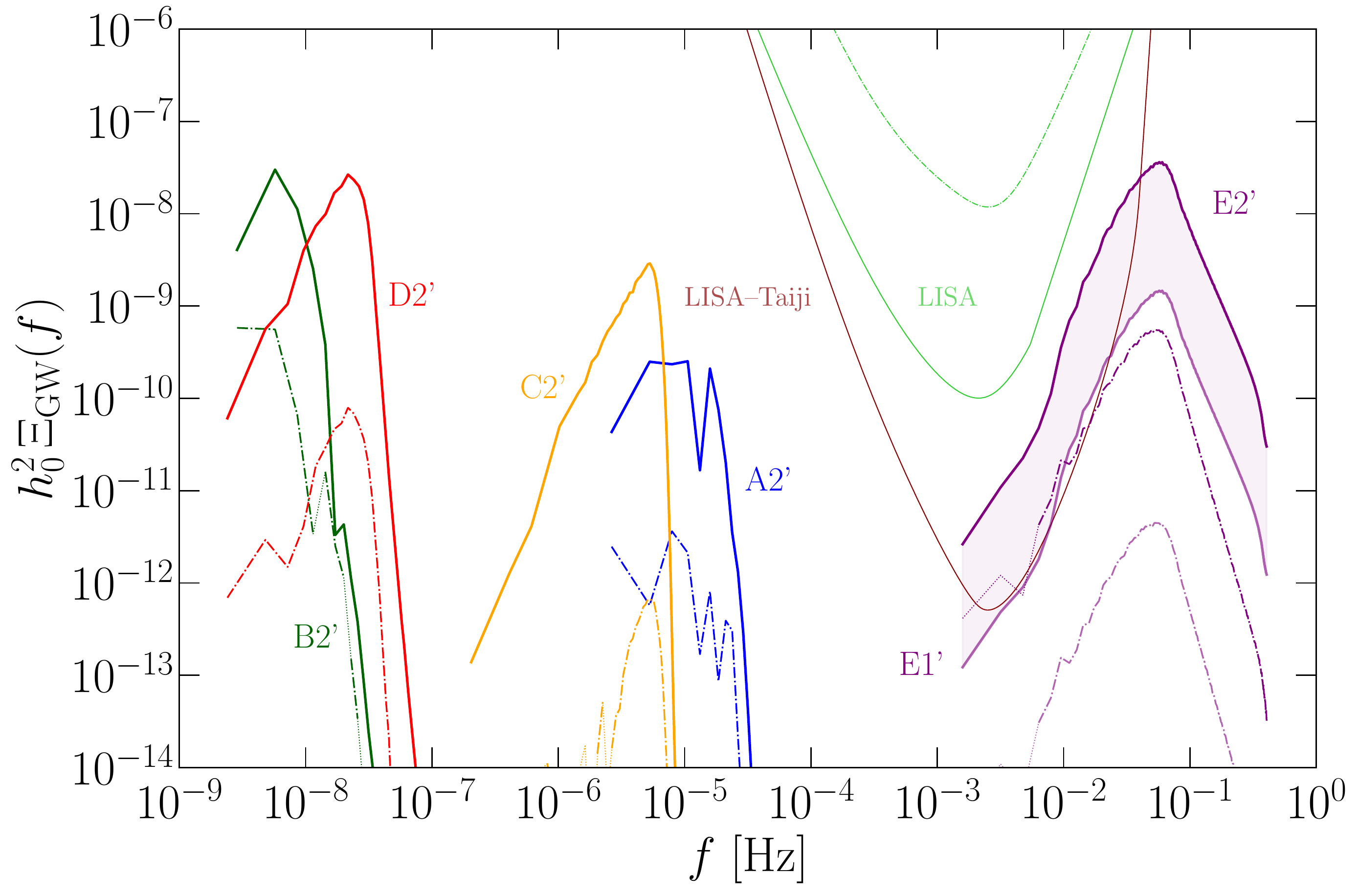}
\caption{
GW energy $h_0^2\OmGW$ (left) and helicity $h_0^2\, \XiGW(f)$ (right) spectra scaled to the present day, in units of physical frequency $f$: 
run A2' is in blue, B2' in green, C2' in orange, D2' in red, and E1' and E2' in purple. 
Solid curves are linear solutions and dashed-dotted curves
are the differences between nonlinear and linear solutions.
Dotted curves correspond to negative values of the differences.
We show the $2\sig$ confidence region for the 30-frequency power law fit for a common-process spectrum reported in the NANOGrav 12.5-year dataset (blue shaded wedge)
\cite{NANOGrav:2020bcs},
as well as the expected sensitivity curve for the SKA (black line) \cite{Moore:2014lga}.
LISA instrument (dash-dotted line) \cite{Robson:2018ifk} and power law sensitivity (PLS) (solid line)
\cite{Caprini:2019pxz, Pol:2021uol} are shown in green.
The PLS of DECIGO \cite{Kawamura:2020pcg} and of the $\mu$Ares concept, proposed within the ESA Voyage 2050 programme \cite{Sesana:2019vho},
are shown in dark red and cyan, respectively, assuming 4 years of observations.
The PLSs estimated for astrometry methods such as Gaia and Theia \cite{Garcia-Bellido:2021zgu}; for atomic interferometer projects AION ($2\km$ design)
\cite{Badurina:2019hst} and AEDGE \cite{AEDGE:2019nxb}, and using
binary resonance forecasts by 2038 from binary
millisecond pulsars (MSPs) and lunar laser-ranging (LRR) measurements
are shown in dark blue, brown, and violet, respectively.
Regarding the detectability of GW helicity (right panel), we show the PLS of LISA
to a polarized GWB using the dipole response function induced by our proper motion
\cite{Domcke:2019zls,Ellis:2020uid,Pol:2021uol} and the one
obtained using the LISA--Taiji network \cite{Orlando:2020oko,Seto:2020zxw,Pol:2021uol}.
All PLS curves assume an SNR of 10.
}
\label{fig:comp2}
\end{figure*}

Table~\ref{tab:runs} and Fig.~\ref{fig:spec_diffA} have demonstrated that the nonlinear effects are subdominant, 
and their spectra at the end of reheating are virtually indistinguishable from the linear ones when $\EEEM$ is smaller than unity.
This is still true when we consider the spectra at later times, with the exception
of the appearance of a bulge below the spectral peaks in the cases with $\beta = 7.3$.
However, these features appear around values much smaller than those near the
spectral peak, making them difficult to be observed in the future.
We explore whether it is sensible to consider the detection prospects of such nonlinear features, as well as the detectability of the signals considered
in the present work.
To do this, we convert the
direct simulation results into present day observable GW energy density $\OmGW$ and helicity $\XiGW$ spectra in terms of physical
frequency $f$, discussed in Sec.~\ref{ssec:EMGW_early_universe}.
Figure~\ref{fig:comp2} presents such estimated GW energy densities
$h_0^2\OmGW(f)$ and $h_0^2\XiGW(f)$ from runs
A2', B2', C2', D2', and E2', listed in \Tab{tab:runs},
which correspond to the upper bound $\EEEM = 0.1$ imposed by the BBN 
limit \cite{Shvartsman:1969mm,Grasso:1996kk,Kahniashvili:2009qi},
and E1', with smaller $\EEEM = 0.02$.
We compare our numerical results with the sensitivities of some current and planned GW detectors:
space-based GW detectors, like the Laser Interferometer Space Antenna 
(LISA) \cite{LISA:2017pwj},
the DECi-hertz Interferometer Gravitational wave Observatory (DECIGO) \cite{Seto:2001qf}, and the $\mu$Ares concept proposed within the ESA
Voyage 2050 programme \cite{Sesana:2019vho};
PTAs, like the North American Nanohertz Observatory for Gravitational Waves (NANOGrav) \cite{NANOGrav:2020bcs}
and the Square Kilometer Array (SKA) sensitivity \cite{Moore:2014lga}; proposed astrometry methods
using Gaia \cite{Moore:2017ity} and Theia \cite{Theia:2017xtk}; atomic interferometer designs,
like the terrestrial Atom Interferometer Observatory and
Network (AION) \cite{Badurina:2019hst} and the 
space-based Atomic Experiment for Dark Matter and Gravity
Exploration (AEDGE) \cite{AEDGE:2019nxb}, and the proposed detections
based on binary resonance \cite{Blas:2021mpc, Blas:2021mqw}.

For both helical and nonhelical magnetogeneses, $\beta=2.7$ (with $\Tr \approx 100 \MeV$) can produce
GWs within the detection range of PTAs,
or astrometry methods using Gaia or Theia.
However, since a primordial GWB is not a transient event and will not
allow detectors to compare linear and nonlinear signals,
very detailed spectral templates might be needed in order to extract specific features regarding nonlinear GWs in future PTA data.
The helical run D2' produces a linear GW solution that would be compatible
with the observations reported by NANOGrav, while the helical run B2'
would lead to a larger GW signal, which has not been observed by current
PTA measurements.
Therefore, $\EEEM < 0.1$ is required for nonhelical signals with an end-of-reheating temperature around QCD.
In both cases, with the improvement of PTA measurements in the coming years, as well as astrometry methods, we might be able to detect the
corresponding nonlinear GW signals.

On the other hand, the runs with $\beta = 7.3$, corresponding to $\Tr = 100$ and
$8 \GeV$ for the nonhelical and helical cases, respectively, peak around
$\sim \! 10^{-5} \Hz$, which is located in between the frequencies reached by PTA
experiments and space-based GW detectors.
There are some proposals to detect GWs in this range of frequencies,
like methods based on binary resonance, which have already been used to
put some upper constraints on GWBs \cite{Blas:2021mpc, Blas:2021mqw},
and next-generation of space-borne GW detectors, like the concept proposed within the ESA Voyage 2050 programme, $\mu$-Ares \cite{Sesana:2019vho}.
The former method, using forecasts of lunar-laser ranging data available by 2038 (see \Fig{fig:comp2}) would allow to probe GWs at $\lesssim$ 1 $\mu$Hz.
However, the resulting signals from A2' and C2' runs are peaking at
10 and 5 $\mu$Hz, respectively.
The expected sensitivity of $\mu$Ares would allow to detect this type
of signals with very large signal-to-noise ratios (SNRs), as well as their nonlinear contribution (especially for the nonhelical run A2').

At larger energy scales, with $\Tr = 3 \times 10^5 \GeV$, the linearized contribution to 
the GW signal of the run E2', with $\EEEM = 0.1$, can be detectable by LISA with a large
SNR $\sim 9500$, peaking at $6 \times 10^{-2} \Hz$.
In addition, the memory effect can also be detectable with a
SNR $\sim 580$.
For lower EM energy density $\EEEM = 0.02$ (E1') we still find
a detectable signal, with SNRs of 380 and 4 for the linear and nonlinear
contributions, respectively.
Next-generation planned space-based GW detectors like DECIGO and Big Bang Observer (BBO)
\cite{Crowder:2005nr} (the latter is not shown in \Fig{fig:comp2} but
it is expected to have a slightly lower sensitivity curve than DECIGO in
a similar range of frequencies) will improve even more the detectability of this
type of signals and their memory effect contributions.
In addition, atomic interferometry projects, like AION and AEDGE, will
be able to detect the GW signal of E2'.
While the former will cover only the high-frequency range of the GW signal
and its nonlinear contribution, the sensitivity of the latter will allow to probe
the full signal.

Alternatively, the polarization features shown in Fig.~\ref{fig:spec_pol}
might offer a different approach for observation.
While PTAs are not capable of detecting circular polarization of GWBs \cite{Belgacem:2020nda}, different methods to detect polarization
have been proposed using space-based GW detectors like LISA and Taiji \cite{Ruan:2018tsw}.
For example, the anisotropies induced in a polarized isotropic
GWB by our proper motion produce a dipolar response
to the GW signal, which allows to detect their polarization \cite{Seto:2006hf,Seto:2006dz},
as studied in Ref.~\cite{Domcke:2019zls} for LISA and in Refs.~\cite{Seto:2020zxw, Orlando:2020oko} for the LISA--Taiji network.
The detectability of polarized GWBs produced by primordial helical
magnetic fields originated or present at the EW scale, following this approach,
has been recently studied in Ref.~\cite{Pol:2021uol}.
We show in \Fig{fig:comp2} the helical GW spectra produced by our runs
and compare them to the sensitivities of LISA (via the dipole response)
and the LISA--Taiji network.
Only the runs of series E' peak
within the range of frequencies where LISA can detect GW signals.
According to Ref.~\cite{Pol:2021uol}, the SNR of 9500, which has been obtained
for run E2', corresponds to an SNR of a fully polarized GWB of $16$ assuming a
flat spectrum.
In our case, this run has an SNR of 8,
which is slightly below the commonly used value of 10 to claim the signal to be
detectable \cite{Caprini:2019pxz}.
The difference in helicity due to the memory effect corresponds to a signal
(see \Fig{fig:comp2}) with an SNR of $\sim\!0.25$.
By considering the LISA--Taiji network, the polarized SNR of a flat spectrum 
would be 3000 following Ref.~\cite{Pol:2021uol}.
In our specific case, the spectral shape of run E2' yields a
smaller value of 360.
The difference in helicity corresponds to an SNR of 12.
For the run E1', with $\EEEM = 0.02$, we find SNRs of 18 and 0.18
for the linear and nonlinear contributions, respectively.
Hence, the polarization spectra corresponding to the series E',
with EM energies between 0.02 and 0.1, could be detectable, and the
nonlinear contribution could be detectable around the $\EEEM = 0.1$
case.

The amplitude of the polarization spectrum decreases when we include the leading-order
nonlinear term in the GW production, as shown in \Fig{fig:spec_pol} and quantified,
using the total polarization, by the suppression coefficient
$r$ (see \Tab{tab:q_p_r} and bottom panel of \Fig{fig:comp1}).
However, this suppression is only significant for large values of the EM energy
density $\EEEM > 1$.
Hence, the memory effect does not affect the prospects of polarization
detectability for $\EEEM \leq 0.1$, which is the upper bound imposed
by the BBN limit.

\section{Conclusions}
\label{sec:conclusions}

In this work, we take the GW equation beyond the standard linear order and study the 
consequent spectral features due to such nonlinearities in the context of
cosmological GWBs driven by inflationary-generated primordial
magnetic fields during reheating era.
Instead of providing a stationary piece to the final strain for
localized transient events such as binary black hole mergers,
the displacement memory effect in the context of a primordial continuous source 
manifests in an overall boost in the GW energy and helicity.
In terms of energy spectra $\EGW(k)$, the nonlinear boosts closely follow
the spectral shapes of the linear solutions, but become more evident by
bulging on top of the linear solutions, when the latter start decreasing
sharply.
The bulges are also extended over a longer range of $k$ in the presence of helicity.
We find that the boost of GW energy density due to the inclusion of the leading-order 
nonlinear term occurs at all wave numbers of the spectrum and we predict
analytically that it is due to terms of order ${\cal O} (h^3)$; see \Eqs{oom_EEGW}{EGWnlin_EGW}.
However, we find a small shift in the phase of the GW mode oscillations due to
the nonlinear term.
Hence, at some specific times, this shift can compensate the nonlinear
boost and smaller values of the nonlinear solution can be found.

Another nonlinear aspect related to helicity can be seen in the polarization spectra
$\PPP_\GW(k)$ where, for $\beta=7.3$, the polarization is weakened approximately at the $k$
range where the bulges occur. 
However, this weakening is not visible for $\beta=2.7$, until the unrealistically large $\EEEM=10$ is reached.
Similarly, in the case with $\beta = 1.7$, we find a slight decrease in the spectral
polarization, which is only significant in the case with $\EEEM = 1$.
In general, we find that the addition of the leading-order nonlinear term 
suppresses polarization at all wave numbers and we show analytically
that this decrease depends on the polarization of the linearized solution and that it is
proportional to terms of the order ${\cal O} (h)$.
This occurs because the nonlinear boost is, in general, stronger in the
GW energy density than in its helicity.

In terms of the total energy, the relation $\Delta\EEGW=(p\EEEM)^3$ is found.
Together with the quadratic relation $\EEGW=(q\EEEM)^2$, found in earlier similar numerical computations,
the coefficients $q$ and $p$ are determined numerically and given in \Tab{tab:q_p_r}.
Furthermore, the scaled coefficients $\tilde{p}\equiv k_*p$ and
$\tilde{q}\equiv k_*q$ are
found to be a better parameterization, presenting less variation over different 
scenarios, which is consistent with the scaling $\EEGW = (q \EEEM/k_*)^2$, also obtained
in previous numerical computations.
Similarly, regarding the total GW polarization,
we find the linear relation $\Delta \PPGW = r \PPGW \EEEM$, where
we define the polarization suppression coefficients $r$, given in \Tab{tab:q_p_r}.

Finally, the detection prospects of such nonlinear GWs are discussed.
We find that PTAs and astrometry methods are relatively promising
in detecting certain low end-of-reheating temperature magnetogenesis-induced GWs.
However, it would be rather challenging to detect nonlinear features
directly, since they correspond to a second-order addition to the linear GWBs.
Hence, very accurate models for the linear contribution are
needed in order to remove it and recover the nonlinear part.
The signals produced considering end-of-reheating temperatures between 8 and $100 \GeV$ are the most challenging to detect, due to the gap of proposed GW
detectors around the $\mu$Hz band.
The $\mu$Ares concept proposed within the ESA Voyage 2050 programme \cite{Sesana:2019vho} could allow
to detect this type of signals.
Space-borne GW detectors like LISA, Taiji, and DECIGO,
and atomic interferometers,
also seem promising candidates to detect higher temperature inflationary
scenarios with expected large SNR for the signals produced by helical
magnetic fields.
In addition, the combination of a network of space detectors, e.g., LISA and Taiji, can provide measurements of the polarization of such signals.
Again, the specific features of the nonlinear contribution are of second-order,
so very accurate models and predictions of the linear counterparts are required.
Overall, further studies are required for a more detailed understanding
of nonlinear GWs themselves, as well as the parameter space of various
GW generation processes.

\section*{Data availability}
The source code used for the simulations of this study,
the {\sc Pencil Code}, is freely available \cite{pencil}.
The simulation data are also available at Ref.~\cite{DATA}.
The calculations, the simulation data, and the routines generating the
plots are also available at \url{https://github.com/AlbertoRoper/GW_turbulence/tree/master/memory_effect}.

\medskip
\begin{acknowledgments}
Support through the grant 2019-04234 from the Swedish Research Council
(Vetenskapsr{\aa}det) and the Shota Rustaveli
National Science Foundation (SRNSF) of Georgia (grant FR/18-1462)
are gratefully acknowledged.
ARP is supported by the French National
Research Agency (ANR) project MMUniverse (ANR-19-CE31-0020).
We acknowledge the allocation of computing resources provided by the
Swedish National Infrastructure for Computing (SNIC)
at the PDC Center for High Performance Computing Stockholm
and the A9 allocation at the Occigen supercomputer
provided by the {\em Grand \'Equipement National de Calcul 
Intensif} (GENCI) to the project ``Opening new windows on Early Universe with multi-messenger astronomy''.

\end{acknowledgments}

%\appendix

\bibliographystyle{apsrev4-1}
\bibliography{references}

\end{document}